\documentclass[twocolumn,showpacs,aps,pra,amsmath,amssymb,superscriptaddress]{revtex4-2}

\usepackage{graphicx}
\usepackage{dcolumn}
\usepackage{bm}
\usepackage{hyperref}
\usepackage{subfigure}

\usepackage{epstopdf}

\usepackage{color,bbm}

\newcommand{\beq}{\begin{equation}}
\newcommand{\eeq}{\end{equation}}
\newcommand{\bqa}{\begin{eqnarray}}
\newcommand{\eqa}{\end{eqnarray}}
\newcommand{\nn}{\nonumber}

\newcommand{\rt}[1]{\sqrt{#1}\,}

\newcommand{\smallfrac}[2]{\mbox{$\frac{#1}{#2}$}}

\newcommand{\half}{\smallfrac{1}{2}}

\newcommand{\sq}[1]{\left[ {#1} \right]}

\newcommand{\tr}[1]{{\rm Tr}\sq{ {#1} }}

\newcommand{\id}{\mathbbm{1}}
\newcommand{\B}{{\mathcal B}}
\newcommand{\str}{\mathcal{S}}
\newcommand{\R}{\mathcal R}

\newcommand{\p}[1]{P_{\bm{#1}}}

\newcommand{\blk}{\color{black}}
\newcommand{\blu}{\color{blue}}
\definecolor{maroon}{rgb}{0.7,0,0}

\definecolor{ngreen}{rgb}{0.3,0.7,0.3}

\definecolor{golden}{rgb}{0.8,0.6,0.1}

\begin{document}

\title{Recycling qubits for the generation of Bell nonlocality between independent sequential observers}

\author{Shuming Cheng}
\affiliation{The Department of Control Science and Engineering, Tongji University, Shanghai 201804, China}
\affiliation{Shanghai Institute of Intelligent Science and Technology, Tongji University, Shanghai 201804, China}
\affiliation{Institute for Advanced Study, Tongji University, Shanghai, 200092, China}

\author{Lijun Liu}
\affiliation{College of Mathematics and Computer Science, Shanxi Normal University, Linfen 041000, China}

\author{Travis J. Baker}
\affiliation{Centre for Quantum Computation and Communication Technology (Australian Research Council),
	Centre for Quantum Dynamics, Griffith University, Brisbane, QLD 4111, Australia}

\author{Michael J. W. Hall}

\affiliation{Department of Theoretical Physics, Research School of Physics, Australian National University, Canberra ACT 0200, Australia}

\date{\today}

\begin{abstract}

There is currently much interest in the recycling of entangled systems, for use in quantum information protocols by sequential observers. In this work, we study the sequential generation of Bell nonlocality via recycling one or both components of two-qubit states. We first give a description of two-valued qubit measurements in terms of measurement bias, strength, and reversibility, and derive useful tradeoff relations between them. Then, we derive one-sided monogamy relations for unbiased observables, that strengthen the recent Conjecture in [S. Cheng {\it et al.}, Phys. Rev. A \textbf{104}, L060201 (2021) ] that if the first pair of observers violate Bell nonlocality then a subsequent independent pair cannot, and give semi-analytic results for the best possible monogamy relation.  We also extend the construction in [P. J. Brown and R. Colbeck, Phys. Rev. Lett. \textbf{125}, 090401 (2020)] to obtain (i)~a broader class of two-qubit states that allow the recycling of one qubit by a given number of observers on one side, and (ii)~a scheme for generating Bell nonlocality between arbitrarily many independent observers on each side, via the two-sided recycling of multiqubit states. Our results are based on a formalism that is applicable to more general problems in recycling entanglement, and hence is expected to aid progress in this field.  

\end{abstract}


\maketitle

\section{Introduction}\label{Sec1: Introduction}

Quantum entanglement is not only fundamental to understanding quantum mechanics, but also is an indispensable resource in various information tasks, such as quantum
teleportation~\cite{Bennett93} and secure quantum key distribution~\cite{Ekert91}. Hence, it is of importance to study how to efficiently use this entanglement resource.  Recently, the possibility that entanglement from the same source can be recycled multiple times, by sequential pairs of independent observers, has been shown by Silva {\it et al.}~\cite{Silva15}.  This has attracted great interest both  theoretically~\cite{Mal16,Curchod17,Tavakoli18,Bera18,Sasmal18,Shenoy19,Das19,Saha19,Kumari19,Brown20,Maity20,Bowles20,Roy20,Zhang21,Cheng21} and experimentally~\cite{Schiavon17,Hu18,Choi20,Foletto20,Feng20,Foletto21,Jie21}. 

\begin{figure*}[!t]
	\centering
	
	\includegraphics[width=
	\textwidth]{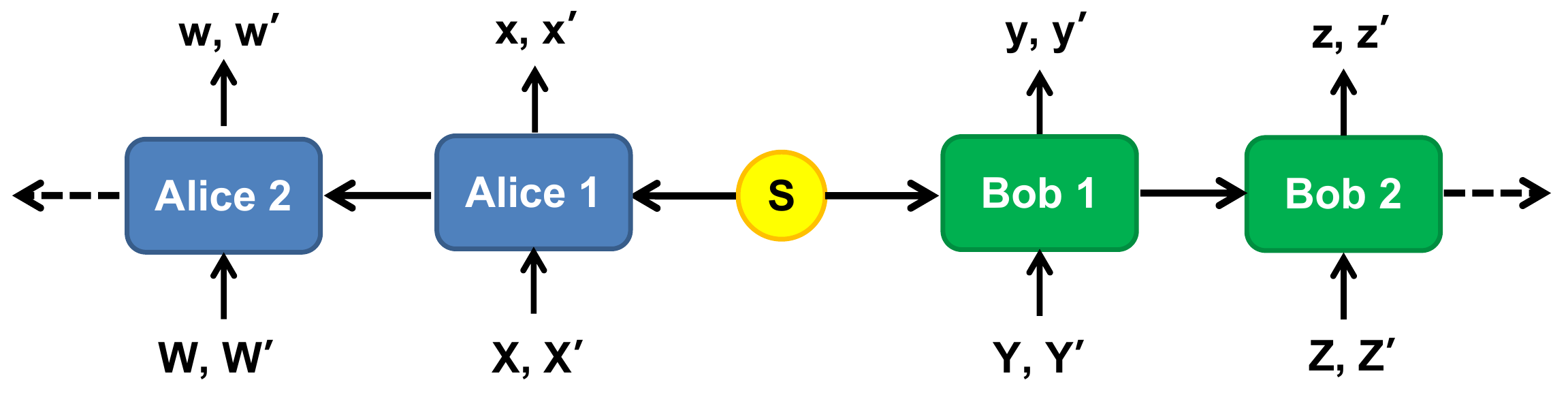}
	
	\caption{Sequential sharing with multiple observers on each side. A source S generates two qubits on each run, which are received by observers Alice~1 and Bob~1 ($A_1$ and $B_1$ in the main text). They each make one of two local measurements on their qubit with equal probabilities; record their result; and pass their qubit onto independent observers Alice~2 and Bob~2, respectively ($A_2$ and $B_2$ in the main text). It is known that Alice~1 can demonstrate Bell nonlocality with each of an arbitrary number of Bobs in this way, by recycling the second qubit~\cite{Brown20}. However, we have recently given strong analytic and numerical evidence for the conjecture that Alice~1 and Bob~1 can demonstrate Bell nonlocality in this manner only if Alice~2 and Bob~2 cannot, and vice versa~\cite{Cheng21}. A similar result holds for the pairs (Alice~1, Bob~2) and (Alice~2, Bob~1)).
	}
	\label{fig:fig1}
\end{figure*}

As illustrated in Fig.~\ref{fig:fig1}, recycling entangled systems typically requires a first pair of observers, Alice~1 and Bob~1, passing their measured systems onto a second pair of observers, Alice~2 and Bob~2, who then pass them onto a third pair, etc. In this scenario, sufficient entanglement can remain, following each measurement, to allow multiple pairs of observers to sequentially implement quantum information protocols such as quantum key distribution~\cite{Ekert91,Ekert14} and randomness generation~\cite{Pironio10,Foletto21}.

In this work, we consider the problem of recycling entangled systems to generate sequential sharing of Bell nonlocality~\cite{Silva15}. For example, if an observer $A$ on one side chooses between two-valued measurements $X$ or $X'$ at random, and an observer $B$ on the other side similarly chooses between $Y$ or $Y'$, then Bell nonlocality can be revealed from the joint measurement statistics via the Clauser-Horne-Shimony-Holt (CHSH) parameter~\cite{Clauser69,Brunner14}
\beq \label{chsh}
S(A,B):= \langle XY\rangle+\langle XY'\rangle+\langle X'Y\rangle - \langle X'Y'\rangle 
\eeq
with $\langle XY\rangle=\sum_{x, y} x y\, p(x, y|X,Y)$, where $x, y\in \{-1, 1\}$ label the corresponding outcomes of measurements $X, Y$. Violation of the Bell-CHSH inequality $S(A,B)\leq2$ certifies the sharing of Bell nonlocality between these two observers. 

Notably, it was shown by Brown and Colbeck that, given a pair of entangled qubits, a single observer is able to share Bell nonlocality with each one of an arbitrarily long sequence of independent observers on the other side~\cite{Brown20}. Surprisingly, however, we have recently found strong analytic and numerical evidence that, under the same assumptions considered in~\cite{Brown20}, it is impossible to recycle {\it both} qubits such that Bell nonlocality is shared between sequential pairs of observers on each side~\cite{Cheng21}. In particular, the evidence supports the conjecture that  observers Alice~1 and Bob~1 in Fig.~\ref{fig:fig1} can violate a CHSH inequality only if Alice~2 and Bob~2 cannot, and similarly for the pairs (Alice~1, Bob~2) and (Alice~2, Bob~1). This restriction of qubit recycling to one side may be viewed as a type of sharing monogamy, and we have given corresponding one-sided monogamy relations that are valid for large classes of states and measurements~\cite{Cheng21}.

In this paper we continue to investigate the sequential sharing of Bell nonlocality via recycling the components of entangled systems, and in particular generalise several results in~\cite{Brown20} and~\cite{Cheng21}. We start with a brief review in Sec.~\ref{Sec2. Qunit Observables} on the characterisation of general two-valued qubit observables introduced in~\cite{Cheng21}, and give a  measurement model that provides a simple interpretation of the strength and bias of such observables. In Sec.~\ref{Sec4. Measurements Reversibility}, we give a general formalism for describing sequential scenarios, based on quantum instruments~\cite{Davies70,Ozawa84}, and review the optimal reversibility properties of square-root measurements in this context. We also give natural definitions of the maximum reversibility and minimum decoherence of a qubit observable; tradeoff relations between these quantities and the strength and bias of the observable; and connections with the class of weak measurements considered by Silva {\it et al.}~\cite{Silva15}.  

The above results provide the tools needed in Sec.~\ref{sec:monog} for obtaining several one-sided monogamy relations (only proved for a special case in~\cite{Cheng21}), for the sequential generation of Bell nonlocality via measurements of unbiased observables. We also give numerical evidence that even stronger monogamy relations hold for this case, and obtain semi-analytic forms for the best possible such relation. In Sec.~\ref{Sec6. One-sided} we apply our tools to scenarios in which recycling is possible for arbitrary numbers of observers. First, if the source is not restricted to generation and measurement of a single qubit pair we show, by generalising the construction by Brown and Colbeck in~\cite{Brown20}, that Bell nonlocality can be sequentially generated between arbitrarily many observers on each side, via recycling multiqubit states. Second, a different generalisation of the Brown-Colbeck construction yields a larger class of two-qubit states for which a single Alice can share Bell nonlocality with a given number of Bobs. 

Conclusions are given in Sec.~\ref{Sec8. Conclusions}.

\section{Two-valued qubit observables} \label{Sec2. Qunit Observables}

In this section, we recap the description of general two-valued qubit observables given in~\cite{Cheng21} (see also~\cite{Yu2010}), and note several important properties for later use. A simple measurement model for such observables is also noted.

\subsection{Strength and bias}\label{subsec2.1}

We can label the outcomes of a general two-valued observable $X$ by $\pm 1$. It is then described by a positive operator valued measure (POVM) $\{X_+,X_-\}$, with $X_\pm\geq 0$, $X_++X_-=\id$, and probability distribution $p_\pm=\tr{\rho X_\pm}$. The observable is equivalently represented by the operator $X= X_+ - X_-$, where $X_\pm=\half(\id\pm X)$, $-\id\leq X\leq \id$ and $\langle X\rangle=p_+-p_-$. This representation is particularly useful for the purposes of the CHSH parameter~(\ref{chsh}), as the expectation value of the product of two such observables $X$ and $Y$, acting on respective components of a quantum system, is given by
\begin{align}
\langle XY\rangle :&= \sum_{x,y=\pm 1} xy\,p(x,y|X, Y) \nn\\
&= \sum_{x,y=\pm1} xy\langle X_x\otimes Y_y\rangle \nn\\
& = \sum_{x,y=\pm1} xy\left\langle \frac{1+xX}{2}\otimes\frac{1+yY}{2}\right\rangle\nn\\
& = \langle X\otimes Y\rangle .
\label{prodxy}
\end{align}

For a qubit, the operator $X$ can be decomposed as
\beq
X =  \B \id + \str\bm \sigma\cdot\bm x \label{observable}
\eeq
with respect to the Pauli spin operator basis $\boldsymbol{\sigma}\equiv (\sigma_1, \sigma_2, \sigma_3)$. Here $\B$ defines the {\it outcome bias} of the observable; $\str\geq0$ denotes its {\it strength}~\cite{Shenoy19} or sharpness~\cite{Yu2010,Choudhary13,Kunjwal14} (and is also called its information gain~\cite{Silva15});  and $\bm x$  is a unit direction associated with the observable, with $|\bm x|:=(\bm x \cdot \bm x)^{1/2}=1$. 
The requirement $-\id\leq X\leq \id$ is equivalent to the constraint
\beq \label{sbcon}
\str + |\B| \leq 1
\eeq
on strength and bias. 

It follows for the case of maximum strength, $\str=1$, that the bias  must vanish, i.e., $X=\bm\sigma\cdot\bm x$ is the projective observable corresponding to spin in direction $\bm x$. Conversely, a minimum strength, $\str=0$, corresponds to the trivial observable $X=\B\id$, equivalent to tossing a two-sided coin with biased outcome probabilities $p_\pm=\half(1\pm\B)$ and average outcome $\langle X\rangle=\B$. 

For later purposes it is useful to note that any two-qubit state $\rho$ can be parameterised in the compact form 
\beq \label{bloch}
\rho=\frac14\sum_{\mu,\nu=0}^4 \Theta_{\mu\nu} \, \sigma_\mu\otimes\sigma_\nu,~~\Theta:=\begin{pmatrix} 1 & \bm b\\ \bm a^\top & T \end{pmatrix} ,
\eeq
with $\sigma_0=\id$. Here $\bm a:=\langle \bm \sigma\otimes \id\rangle$ and $\bm b:=\langle \id\otimes \bm \sigma\rangle$ refer to Alice's and Bob's Bloch vectors respectively, and $T:=\langle \bm\sigma\otimes\bm\sigma^\top\rangle$ is the spin correlation matrix. Consequently,  Eq.~(\ref{prodxy}) can be rewritten as
\beq \label{prodxy2}
\langle XY\rangle = \begin{pmatrix} \B_X&  \str_X \bm x^\top\end{pmatrix}
\begin{pmatrix} 1 & \bm b^\top\\ \bm a & T \end{pmatrix} 
\begin{pmatrix} \B_Y\\ \str_Y \bm y \end{pmatrix} ,
\eeq
for $X=\B_X\id+\str_X\bm\sigma\cdot\bm x$ and $Y=\B_Y\id+\str_Y\bm\sigma\cdot\bm y$. In the case of projective observables $X=\bm \sigma\cdot \bm x$ and $Y=\bm\sigma\cdot \bm y$, the right hand side simplifies to the familiar expression $\bm x^\top T\bm y$. In contrast, for trivial observables $X=Y=\id$ the right hand side simplifies to 1, implying that  a corresponding CHSH parameter of $S(A,B)=2$ in Eq.~(\ref{chsh}) can be always be obtained via trivial observables.

\subsection{A simple measurement model}\label{subsec2.2}

A simple interpretation of a general qubit observable $X$ as a noisy projective observable, with the strength, bias and post-measurement state having correspondingly simple interpretations, is as follows.  

In particular, suppose that one either (i)~measures the projective observable $\bm\sigma\cdot\bm x$, with a `success' probability $\str$, or (ii) otherwise assigns outcomes $\pm 1$ by flipping a coin having biased outcome probabilities $q_\pm=\half(1\pm\epsilon)$, with probability $1-\str$. The resulting measurement statistics  are therefore generated by the POVM elements
\beq
X_\pm = \str \frac{\id\pm\bm\sigma\cdot\bm x}{2} + (1-\str)\frac{1\pm\epsilon}{2}\id .
	\eeq
This corresponds to the observable
\beq
X = X_+-X_- = \epsilon (1-\str)\id + \str\bm\sigma\cdot\bm x ,
\eeq
having strength $\str$ and bias $\B=\epsilon(1-\str)$. Note that constraint~(\ref{sbcon}) is equivalent to the property $|\epsilon|\leq1$.

There are many different ways to measure a given observable, and the post-measurement state depends on the measurement details (see Sec.~\ref{subsec4.1}). However, it is of interest to consider the post-measurement state for the simple implementation above if  the projective measurement  is assumed to leave the qubit in the corresponding eigenstate of $\bm \sigma\cdot\bm x$, while the coin flip leaves the qubit unchanged. It follows that if the state prior to the measurement is described by density operator $\rho$, then the post-measurement state is described by
\begin{align} \label{simplemeas}
\rho_{\str}&=\str\left( \p{x}\rho\p{x}+\p{-x}\rho\p{-x}\right) + (1-\str)\rho \nn\\
&= \p{x}\rho\p{x}+\p{-x}\rho\p{-x} +(1-\str)\left(\p{x}\rho\p{-x}+\p{-x}\rho\p{x}\right)
\end{align} 
where $\p{x}:=\half(\id+\bm\sigma\cdot\bm x)$. 
Thus, the diagonal elements of the state with respect to the $\bm\sigma\cdot\bm x$ basis are unchanged, while the off-diagonal elements are scaled by a factor $1-\str$, implying that the latter provides a measure of the reversibility of the measurement. We will see in Sec.~\ref{subsec4.3}, however, that measurement implementations having larger degrees of reversibility are possible for all $0<\str<1$.

	\section{Measurements and reversibility} \label{Sec4. Measurements Reversibility}

In the context of sequential generation of entanglement, as depicted  in Fig.~\ref{fig:fig1}, the effect of a measurement on the subsequent state of a quantum system is critical.
In this section we first consider the general form of the post-measurement states, and then focus on the special case of square-root measurements. The latter have been previously argued to correspond to the maximally reversible measurements of any given observable. For two-valued qubit observables this leads to natural measures of maximum reversibility and minimum decoherence, which are completely determined by the strength and bias of the observable. 

\subsection{General considerations}\label{subsec4.1}

We briefly review measurements on general quantum systems here, and specialise to the case of qubit systems in the following subsections.
 
A measurement on a  quantum system described by density operator $\rho$, with outcome $x$, will leave the system in some corresponding state $\rho_x$ with probability $p_x$ (we do not restrict to $x=\pm1$ here). The most general description of such a measurement is an {\it instrument}~\cite{Davies70,Ozawa84}, i.e., a set of completely positive (CP) maps,  $\{\phi_x\}$, satisfying
$\phi_x(\rho) = p_x \rho_x$.
Taking the trace of each side then yields
\beq
p_x = \tr{\phi_x(\rho) }=\tr{\id\phi_x(\rho)} = \tr{\phi^*_x(\id)\rho} ,
\eeq
where $\chi^*$ denotes the dual map of  linear map $\chi$ (defined via $\tr{A\chi(B)}= \tr{\chi^*(A)B}$ for all $A,B$).  It follows that the observable measured by the instrument is described by the POVM $\{X_x\}$ with
\beq
X_x := \phi^*_x(\id).
\eeq
Moreover, the density operator describing an ensemble of such systems after measurement is given by the completely-positive trace-preserving (CPTP) map
\beq \label{cptp}
\phi(\rho) := \sum_x p_x \rho_x  = \sum_x \phi_x(\rho).
\eeq
In the simplest case,
\beq \label{kraus}
\phi_x(\rho) = M_x \rho M_x^\dagger,\qquad X_x = M^\dagger_x M_x ,
\eeq
for a set of `measurement operators' $\{M_x\}$, so that each post-measurement state $\rho_x$ is pure when $\rho$ is pure. More generally, however, each $\phi_x(\rho)$  is a sum of such terms.

If a local measurement described by the CPTP map $\phi$ is made on the first component of an ensemble of bipartite quantum systems described by $\rho$, it follows that the post-measurement state of the ensemble is given by $\rho'=(\phi\otimes I)(\rho)$. For the purpose of explicit calculations, it is convenient to choose trace-orthogonal basis sets $\{\tilde\sigma_\alpha\}$ and $\{\tilde\tau_\mu\}$ for the Hermitian operators of the first and second components, respectively, such that $\tr{\tilde\sigma_\alpha\tilde\sigma_\beta}=c\delta_{\alpha\beta}$ and $\tr{\tilde\tau_\mu\tilde\tau_\nu}=d\delta_{\mu\nu}$ for two constants $c$ and $d$. This gives the generalised Bloch representation $\rho=(cd)^{-1}\sum_{\alpha\mu} \tilde\Theta_{\alpha\mu}\tilde\sigma_\alpha\otimes\tilde\sigma_\mu$, with $\tilde\Theta_{\alpha\mu}:=\langle \tilde\sigma_\alpha\otimes\tilde\sigma_\mu\rangle$, generalising Eq.~(\ref{bloch}).  Hence, a local measurement on the first component of the system, taking $\rho$ to $\rho'=(\phi\otimes I)(\rho)$ with $\phi$ as in Eq.~(\ref{cptp}), takes $\tilde\Theta$ to $\tilde\Theta'$ with
\begin{align}
	\tilde\Theta'_{\alpha\mu}
	&= \tr{(\phi\otimes I)(\rho)\,\tilde\sigma_\alpha\otimes\tilde\tau_\mu}\nn\\
	&= \frac{1}{cd} \sum_{\beta,\nu}\tilde\Theta_{\beta\nu} \tr{(\phi\otimes I)(\tilde\sigma_\beta\otimes \tilde\tau_\nu)\,\tilde\sigma_\alpha\otimes\tilde\tau_\mu}\nn\\
	&=\frac{1}{cd} \sum_{\beta,\nu}\tilde\Theta_{\beta\nu} \tr{\phi(\tilde\sigma_\beta)\tilde\sigma_\alpha \otimes \tilde\tau_\nu\tilde\tau_\mu}\nn\\
	&=\frac{1}{cd} \sum_{\beta,\nu}\tilde\Theta_{\beta\nu} \tr{\phi(\tilde\sigma_\beta)\tilde\sigma_\alpha }\,\tr{ \tilde\tau_\nu\tilde\tau_\mu}\nn\\
	&= \frac{1}{c} \sum_\beta \tilde\Theta_{\beta\mu} \tr{\phi(\tilde\sigma_\beta)\tilde\sigma_\alpha }.
	\label{thetaprime}
\end{align} 
Thus,
\beq \label{ktheta}
\tilde\Theta' =  \mathcal{K}\tilde\Theta, \qquad\mathcal{K}_{\alpha\beta}:= c^{-1}\tr{\tilde\sigma_\alpha\phi(\tilde\sigma_\beta)}.
\eeq
and the effect of the measurement on an ensemble corresponds to left multiplication of $\tilde\Theta$ by the matrix $\mathcal{K}$. More generally, if local measurements described by CPTP maps $\phi$ and $\chi$ are made on the first and second components, respectively, then the corresponding state $\rho''=(\phi\otimes\chi)(\rho)$ corresponds to transforming $\tilde\Theta$ to
\beq \label{kthetal}
\tilde\Theta''= \mathcal{K} \tilde\Theta \mathcal{L}^\top,\qquad \mathcal{L}_{\mu\nu}:= d^{-1}\tr{\tilde\tau_\mu\blu \chi \blk (\tilde\tau_\nu)},
\eeq
similarly to Eq.~(\ref{ktheta}) above. This easily generalises to sequences of local measurements on each side (see Sec.~\ref{sec:arbab}).

\subsection{Square-root measurements and\\ maximum reversibility}\label{subsec4.2}

There are many possible ways of measuring a general POVM observable $\{X_x\}$. For example, the square-root measurement corresponds to the instrument $\{\phi_x\}$ defined by $\phi_x(\rho):=X_x^{1/2}\rho X_x^{1/2}$, with corresponding CPTP map
\beq \label{sqrt}
\phi_{1/2}(\rho)=\sum_x\phi_x(\rho)=\sum_x X_x^{1/2}\rho X_x^{1/2}  .
\eeq
This example is of fundamental interest, as {\it any} instrument $\{\phi^G_x\}$ describing a measurement of the POVM $\{X_x\}$ has the form $\phi^G_x=\psi_x\circ\phi_x$, for suitable CPTP maps $\psi_x$~\cite{Brown20}.
Thus, any measurement of $X$ formally corresponds to first carrying out the square-root measurement, and then applying a quantum channel to the state depending on the result obtained.

It follows, noting that $\psi_x$ is reversible if and only if it is unitary, that a general measurement of $X$ can be no more reversible than a square-root measurement of $X$. Moreover, if the outcomes are not known (e.g., to a second observer who receives the qubit in a sequential scenario), then a general measurement can only be `reversed' to the square-root measurement if $\psi_x(\rho) = U\rho \,U^\dagger$ for some unitary transformation $U$. In this sense the square-root measurement is the maximally reversible measurement of $X$, up to a unitary transformation.  

The above property of square-root measurements leads to a natural measure of the maximum reversibility for any measurement of a two-valued qubit observable. In particular, the action of a square-root measurement of $X\equiv\{X_+,X_-\}$ on a qubit in state $\rho$  may be calculated from Eqs.~(\ref{observable}) and~(\ref{sqrt}) as~\cite{Cheng21}
\beq
\phi_{1/2}(\rho)  = \p{x}\rho\p{x} + \p{-x}\rho\p{-x}+ \R\left(\p{x}\rho\p{-x}+\p{-x}\rho\p{x}\right),
\label{rhoprime}
\eeq
where $\p{x}=\half(1+\bm\sigma\cdot\bm x)$ and  the parameter $\R$ is given by
\beq 
\R = \half\sqrt{(1+\B)^2-\str^2} + \half\sqrt{(1-\B)^2-\str^2}, \label{reversibility}
\eeq
Thus, $\R=0$ for projective measurements ($\str=1, \B=0$), and $\R=1$ for trivial measurements ($\str=0$). More generally, $\R$ scales the off-diagonal elements of $\rho$, making it a suitable measure of the reversibility of the square-root measurement. Accordingly, it is also a measure of the maximum reversibility of any measurement of $X$.  We will often just refer to it as the reversibility in what follows. 

The interpretation of $\R$ as a measure of maximum reversibility may also be more directly justified in some cases. For example, Silva {\it et al.} introduced a class of weak measurements of unbiased qubit observables, i.e., with $\B=0$ in Eq.~(\ref{observable}), for which the post-measurement state has the form~\cite{Silva15}
\beq
\rho_F:= \p{x}\rho\p{x} + \p{-x}\rho\p{-x}+ F\left(\p{x}\rho\p{-x}+\p{-x}\rho\p{x}\right),
\label{rhof}
\eeq
where $F$ is a `quality factor' that depends on the properties of the pointer state used in the measurement. Comparing Eqs.~(\ref{rhoprime}) and~(\ref{rhof}), it is seen that $F$ is a measure of the reversibility of such a weak measurement. However, as shown in~\cite{Cheng21}, 
\beq \label{frineq}
F\leq \R,
\eeq
with equality holding for a set of optimal pointer states. Thus the reversiblity of such weak measurements is explicitly bounded above by the maximum reversiblity $\R$.  It will be shown in the next subsection that $\R$ also explicitly upper bounds the reversibility of the simple measurement protocol in Eq.~(\ref{simplemeas}), for both biased and unbiased observables.

Moreover, as noted in~\cite{Cheng21}, if the measurement of a general qubit observable $X=\{X_+,X_-\}$ is implemented with Kraus operators $M_\pm$ as in Eq.~(\ref{kraus}, then the average state disturbance, as quantified by the fidelity $\mathcal{F}$ in~\cite{Banaszek01}, is upper bounded by 
\beq
\mathcal{F}\leq (\R+2)/3,\label{fidelity}
\eeq
where the equality is saturated for the square-root measurement with $M_\pm=X_\pm^{1/2}$. Thus, there is a direct connection between maximum reversibilty and maximum fidelity for this case.

Finally, for the purposes of applying the Horodecki criterion~\cite{Horodecki95}  to the post-measurement state of a two-qubit ensemble, we also need to determine how the spin matrix $T$ transforms under local measurements. For the case of a local measurement on the first qubit, one finds from Eqs.~(\ref{bloch}) and~(\ref{ktheta}) (with $\tilde\sigma_\alpha=\tilde\tau_\alpha=\sigma_\alpha$ and $c=d=2$) that $T$ is mapped to
\beq \label{tprime}
T'= \half\tr{\phi(\id)\bm \sigma} \bm b^\top + KT, 
\eeq
where $K$ is the $3\times3$ matrix with coefficients $K_{jk}=\mathcal{K}_{jk}$.
It follows that $T$ transforms most simply when the first term vanishes, i.e., when (i)~$\bm b=\bm 0$ or (ii)~$\phi(\id)=\id$. 
Note that condition~(ii) is equivalent to the map $\phi$ being unital. 
More generally, if $\phi$ and $\chi$ are unital, or if the local states are maximally mixed, then Eq.~(\ref{kthetal}) yields the simple transformation law
\beq \label{ktl}
T'' = KTL^\top
\eeq
for the spin correlation matrix, following local measurements on each side, with $L_{jk}:=\mathcal{L}_{jk}$.

In particular, noting from Eq.~(\ref{sqrt}) that square-root measurements are unital, i.e., $\phi_{1/2}(\id)=\id$, it follows that local square-root measurements of the qubit observables $X=\B_X\id+\str_X\bm\sigma\cdot\bm x$ and $Y=\B_Y\id+\str_Y\bm\sigma\cdot\bm y$ result in a spin correlation matrix of the form given in Eq.~{\ref{ktl}).  Explicit calculation of $K$ and $L$, from Eq.~(\ref{rhoprime}) and $\mathcal K$ and $\mathcal L$ in Eqs.~(\ref{ktheta}) and~(\ref{kthetal}), then gives~\cite{Cheng21}
\beq \label{txy}
T^{XY}=K^X T K^Y,~~K^X:=\R_XI_3+(1-\R_X)\bm x\bm x^\top,
\eeq
where $I_3$ is the $3\times3$ identity matrix, and $\R_X, \R_Y$ label the maximum reversibilities of observables $X, Y$, respectively. This result will be used in obtaining the one-sided monogamy relations in Sec.~\ref{sec:monog}.

\subsection{Tradeoffs between strength, reversibility\\ and decoherence}\label{subsec4.3}

It has been shown previously that the strength and maximum reversibility of a qubit observable $X=\B\id+\str\bm \sigma\cdot\bm x$ satisfy the tradeoff relation~\cite{Cheng21}
\beq \label{rssum}
\R^2 + \str^2 \leq 1,
\eeq
with equality for the case $\B=0$. Thus, the greater the strength or sharpness of the observable, the less reversibly it can be measured, and vice versa.  
This tradeoff is closely related to the information-disturbance relation of Banaszek~\cite{Banaszek01}, for the case of qubit measurements~\cite{Cheng21}, and will be crucial to the derivation of one-sided monogamy relations in Sec.~\ref{sec:monog}.

Here we  briefly note several further connections between strength, reversibility, bias and information-disturbance; compare the maximum reversibility to that of the simple measurement protocol in Eq.~(\ref{simplemeas});  and  introduce a natural measure of the minimum decoherence of a qubit measurement.

First, we extend tradeoff relation~(\ref{rssum}) to the inequality chain
\beq \label{chain}
1-\str \leq \R^2 \leq 1-\str^2.
\eeq
Thus, a given strength sets both upper and lower bounds on the maximum reversibility. To obtain these bounds, note first from Eq.~(\ref{reversibility}) that $\R=0$ is only possible if $\str=1\pm\B$,  yielding in turn $\B=0$ and $\str=1$,  and that $\R=1$ is only possible if $\str=0$, since otherwise Eq.~(\ref{reversibility}) gives $\R<\half(1+\B)+\half(1-\B)=1$. Hence, Eq.~(\ref{chain}) is certainly valid for  $\R=0$ or 1. Moreover, for $0<\R<1$ it follows directly from Eq.~(\ref{reversibility}) that~\cite{Cheng21}
\beq
\R^2+\str^2 = 1 - \B^2(1/\R^2 - 1),
\eeq
immediately implying the right hand inequality of Eq.~(\ref{chain}). Further, rewriting the above equality as
\beq
|\B|= \R\sqrt{1-\frac{\str^2}{1-\R^2}},
\eeq
and substituting into Eq.~(\ref{reversibility}) gives
\beq
\R = \max\left\{ \R,\sqrt{1-\frac{\str^2}{1-\R^2}}\right\},
\eeq
which immediately implies the left hand inequality of Eq.~(\ref{chain}).

The lower bound in Eq.~(\ref{chain}) has several applications. For example, it implies that the reversibility of the simple measurement protocol in Eq.~(\ref{simplemeas}), i.e., $\R_{\rm simp}=1-\str$, is always upper bounded by the maximum reversibility $\R$. In particular, we have
\beq \label{rsimp}
\R_{\rm simp}=1-\str \leq \sqrt{1-\str} \leq \R ,
\eeq
with strict inequality for $0<\str<1$. This result also implies that the lower bound in Eq.~(\ref{chain}) is stronger than the `disturbance-reversibility' relation given in Theorem~2 of~\cite{Lee20}, for the case of qubit measurements, as the latter relation reduces to  $\R+\str\leq1$ for this case. Lastly, combining constraint~(\ref{sbcon}) with the lower bound in Eq.~(\ref{chain}) gives
\beq
|\B|\leq \R^2,
\eeq
i.e., the outcome bias sets a lower bound on the maximum reversibility of the measurement. 

Noting that the maximum reversibility $\R$ in Eq.~(\ref{rhoprime}) scales the off-diagonal elements of the square-root measurement, it is natural to define a corresponding ``minimal decoherence" by~\cite{Cheng21}
\beq
{\cal D}= \rt{1 - \R^2}. \label{decoherence}
\eeq
Equation~(\ref{chain}) is then equivalent to
\beq
{\cal D} \geq \str \geq {\cal D}^2. \label{decoherence2}
\eeq
In particular, the minimal decoherence of any qubit measurement is at least as large as the strength of the observable being measured. 

Finally, we note that the strength and bias of a given observable can be simply parameterised in terms of its maximum reversibility via
\beq
\str = \sqrt{1-\R^2}\cos\alpha,~~~ \B=\R\sin\alpha,~~~ |\alpha|\leq \sin^{-1}\R,
\eeq
as can be checked by direct substitution into Eq.~(\ref{reversibility}). This parameterisation is useful for numerical searches over general observables, and for deriving further tradeoff relations such as the lower bound
\beq
\R^2+\str^2 =  1-(1-\R^2)\sin^2\alpha\geq 1-(1-\R^2)\R^2\geq\frac34 ,
\eeq
complementary to Eq.~(\ref{rssum}).

\section{One-sided monogamy relations for unbiased observables} 
\label{sec:monog}

\subsection{Overview}
\label{sec:overview}

We now consider the scenario in Fig.~\ref{fig:fig1}, in which observers $A_1$ and $B_1$ make measurements on a pair of entangled qubits, and pass them on to observers $A_2$ and $B_2$, respectively. As discussed in the Introduction, we have previously given strong support for the conjecture that, in the scenario where $A_1$ and $B_1$ each choose between two observables with equal probabilities, they can violate the CHSH inequality if and only if $A_2$ and $B_2$ cannot~\cite{Cheng21}. Here we numerically and analytically investigate this conjecture further, for the particular case of {\it unbiased} observables, including strengthening it and proving several one-sided monogamy relations for this case.

In the light of the discussion in Sec.~\ref{subsec4.2}, we limit our consideration to the scenario where $A_1$ and $B_1$ make square-root measurements of their observables, i.e., to maximally-reversible measurements~\cite{Brown20}. Hence, if $A_1$ measures either $X$ or $X'$ with equal probability, and $B_1$ measures $Y$ or $Y'$ with equal probability, it follows via Eq.~(\ref{txy}) that the spin correlation matrix $T$ of the initial shared state is transformed to $KTL$,
where
\begin{align} \label{kdef}
K&:= \half(K^X+K^{X'}) \nn\\
&~= \frac{\R_X+\R_{X'}}{2}I_3 + \frac{1-\R_X}{2}\bm x\bm x^\top+ \frac{1-\R_{X'}}{2}\bm x'\bm x'^\top 
\end{align}
and
\begin{align} \label{ldef}
L&:= \half(K^Y+K^{Y'}) \nn\\
&~= \frac{\R_Y+\R_{Y'}}{2}I_3 + \frac{1-\R_Y}{2}\bm y\bm y^\top+ \frac{1-\R_{Y'}}{2}\bm y'\bm y'^\top .
\end{align}
This transformation rule allows us to avoid having to explicitly optimise over the set of observables that can be measured by $A_2$ and $B_2$~\cite{Cheng21}. In particular, we can apply the Horodecki criterion to the post-measurement state~\cite{Horodecki95}, to conclude that $A_2$ and $B_2$ can violate the CHSH inequality if and only if
\beq \label{sa2b2}
S^*(A_2,B_2)=2\sqrt{s_1(KTL)^2+s_2(KTL)^2} >2 .
\eeq

In previous work, a search over the possible values of $S(A_1,B_1)$ and $S^*(A_2,B_2)$ strongly supported the conjecture that the pairs $(A_1,B_1)$ and $(A_2,B_2)$ cannot both violate the CHSH inequality~\cite{Cheng21} (see also Fig.~\ref{fig:monog} below). 
It was further shown, analytically, that for the case of unbiased observables $X, X',Y,Y'$ satisfying the assumptions of equal strengths $\str_X=\str_{X'}, \str_Y=\str_{Y'}$ and orthogonal relative angles $\bm x\cdot \bm x'=0=\bm y \cdot \bm y'$ on each side, one has the one-sided monogamy relation~\cite{Cheng21}
\beq \label{monogorthog}
|S(A_1,B_1)| + S^*(A_2,B_2) \leq \frac{8\sqrt{2}}{3} < 4 .
\eeq
It immediately follows from this relation that the quantities $S(A_1,B_1)$ and $S^*(A_2,B_2)$ cannot both be greater than 2, thus proving the conjecture under these assumptions. A numerical search and quadratic one-sided monogamy relations also supported a similar conjecture for the pairs $(A_1,B_2)$ and $(A_2,B_1)$~\cite{Cheng21}, but the latter case will not be considered further here.

In the remainder of this Section we will strengthen the above results in several ways. First, we will give  numerical evidence for the following. 
{\flushleft  \bf Conjecture  for unbiased observables: } {\it For arbitrary unbiased observables $X,X'$ and $Y,Y'$, that are independently measured by $A_1$ and $B_1$ with equal respective probabilities, the one-sided monogamy relation 
\beq \label{unbiasedmonog}
|S(A_1,B_1)| + S^*(A_2,B_2) \leq 4
\eeq 
is always satisfied.}\\
This conjecture immediately implies that the pairs $(A_1,B_1)$ and $(A_2,B_2)$ cannot both violate the CHSH inequality, for any measurements of unbiased observables by the first pair, without any assumptions on their strengths and relative angles. The numerical evidence further shows that stronger one-sided monogamy relations must exist, and we obtain some semi-analytic results for the form of the optimal such relation. 

We will also analytically support the above conjecture, by (i)~proving that the monogamy relation~(\ref{unbiasedmonog}) holds for all unbiased observables with equal strengths on each side (irrespective of the relative angles); and  (ii)~proving that the stronger monogamy relation~(\ref{monogorthog}) holds for all unbiased observables with orthogonal relative angles on each side (irrespective of their strengths).

\begin{figure}[!t]
	\centering
		\includegraphics[width=	0.5\textwidth]{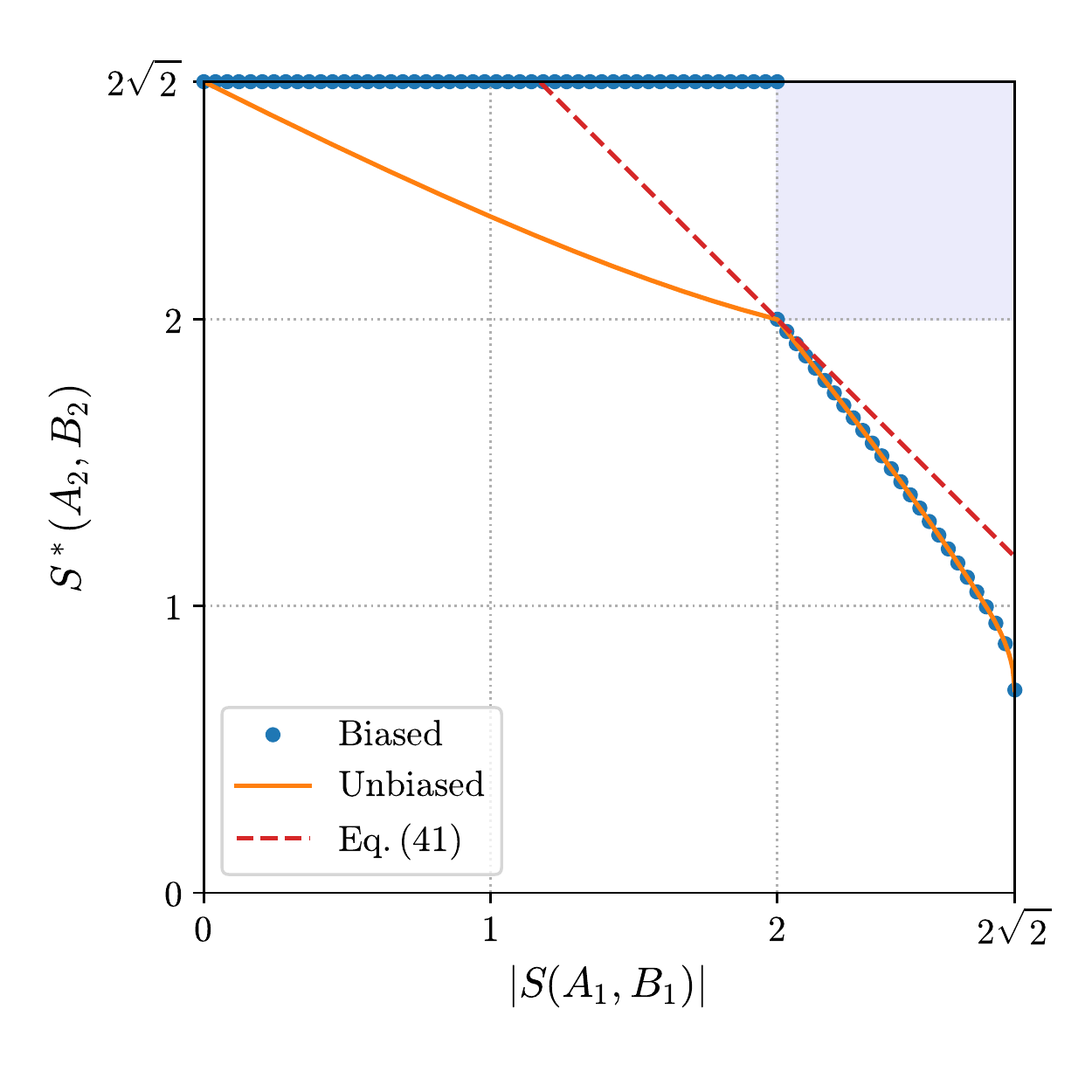}
	
	\caption{ One-sided monogamy relations for biased and unbiased observables. For general biased observables, a global optimisation algorithm was employed in~\cite{Cheng21} to determine the joint range of possible values of $S(A_1, B_1)$ and $S^*(A_2, B_2)$, the results of which are reproduced here as the blue dots.
When restricted to unbiased observables, new numerical results are displayed as the solid orange curve, which lies strictly below the dashed red line corresponding to the conjectured monogamy relation~(\ref{unbiasedmonog}), and which completely coincides with the blue dots for the case $|S(A_1,B_1)|\geq2$. We have further verified that the solid orange curve can be achieved by measuring unbiased observables on a singlet state. The detailed form of the solid orange curve is discussed in Sec.~\ref{sec:semi}.
	}
	\label{fig:monog}\end{figure}

\subsection{Numerical evidence for the conjecture}
\label{sec:num}

The joint range of achievable values of $S(A_1,B_1)$ and $S^*(A_2,B_2)$ can be numerically determined via a global numerical optimisation algorithm, as described in~\cite{Cheng21}. The algorithm searches over all two-valued observables $X, X',Y,Y'$ that can be measured by $A_1$ and $B_1$, and over all pure states (convexity implies that only pure states need be considered), i.e., over a total of 17 free parameters~\cite{Cheng21}. More precisely, a differential evolution optimizer is implemented to seek solutions to the problem
\begin{equation}
\begin{split}
\max\limits_{\alpha,X,X',Y,Y'}  & S^*(A_2,B_2) \\
\text{s.t.} \qquad& S(A_1,B_1) = s,
\end{split}
\label{eq:optimization_problem}
\end{equation}
for fixed values of $s\in[0,2\sqrt{2}]$.
The codes used in the simulations reported here are freely available at~\cite{paper_codes}.

The results of this search are reproduced here as the blue dots in Fig.~\ref{fig:monog}, which plots the numerically-determined maximum value of $S^*(A_2,B_2)$ for each possible value of $S(A_1,B_1)$ (only a subset of points is plotted, for ease of viewing). 
It is seen that $S^*(A_2,B_2)$ can reach the maximum value of $2\sqrt{2}$, and thus allow the pair $(A_2,B_2)$ to violate the CHSH inequality, for any value $|S(A_1,B_1)|\leq 2$ (e.g., via $A_1$ and $B_1$ making non-disturbing trivial  measurements $X=X'=Y=Y'=\B \id$ on a singlet state and $A_2$ and $B_2$ making the optimal CHSH measurements). In contrast, for $|S(A_1,B_1)|>2$, i.e., when the pair $(A_1,B_1)$ can violate the CHSH inequality, the results show that $S^*(A_2,B_2)$ is strictly less than 2. Thus, the dotted blue curve confirms the general conjecture in~\cite{Cheng21} that it is impossible for both pairs to violate the CHSH inequality.

Figure~\ref{fig:monog} also presents new numerical results, for the case where $(A_1,B_1)$ are restricted to measurements of {\it unbiased} observables, corresponding to the solid orange curve. These results were generated by the same method described above but with the biases of the observables set equal to zero. Noting that the dashed red line in Fig.~\ref{fig:monog} corresponds to equality in Eq.~(\ref{unbiasedmonog}), these numerical results therefore strongly support the above conjecture that the one-sided monogamy relation~(\ref{unbiasedmonog}) holds for measurements of unbiased observables. 
We have also numerically verified that for this case the same solid orange curve is obtained under the restriction to a singlet state, in agreement with the reasoning given~\cite{Cheng21}, and that it is also obtained under a further restriction of measurement directions to the equatorial plane.

It is of interest that the dotted blue and solid orange curves are the same, up to numerical error, for any given violation of the CHSH inequality by $A_1$ and $B_1$, i.e., the corresponding maximum possible value of $S^*(A_2,B_2)$ can be achieved even if $A_1$ and $B_1$ are restricted to measure unbiased observables.

\subsection{Semi-analytic optimal monogamy relations}
\label{sec:semi}

The numerical results depicted in Fig.~\ref{fig:monog} support the conjectured one-sided monogamy relation in Eq.~(\ref{unbiasedmonog}), but they also indicate that this relation is not optimal. In particular, an optimal monogamy relation for unbiased observables would reproduce the numerically-generated orange boundary curve in Fig~\ref{fig:monog}. It is therefore of interest to probe the numerical results more closely, to gain information about the possible analytic form of this curve. Some success in this direction is achieved below for several portions of the boundary curve, and we refer to the results, guided by both numerical and analytic analysis, as `semi-analytic' monogamy relations. Strictly analytic but less general relations will be derived in Sec.~\ref{sec:analytic}.

To find suitable ansatzes for the optimal measurement strengths and directions that generate the orange boundary curve in Fig.~\ref{fig:monog}, we begin from the observation in Sec.~\ref{sec:num} that the same curve is numerically generated under the restrictions that (i) the initially-shared state is a singlet state ($T=-I_3$) and (ii) $A_1$ and $B_1$'s observables are confined to the equatorial plane of the Bloch sphere (henceforth setting the $z$-component of all measurement Bloch vectors to zero). Noting that the singlet state is rotationally invariant, we choose $\bm x = (0,1,0)^\top$ without loss of generality.

The numerical results indicate that the optimal measurement parameters under the above restrictions take different forms in three piecewise regions of the orange boundary curve in Fig.~\ref{fig:monog}, given by $|S(A_1,B_1)|\leq 2$, $2<|S(A_1,B_1)| \lesssim 2.72$, and $|S(A_1,B_1)|\gtrsim 2.72$. These regions are therefore considered in turn below.

\begin{figure}[!t]
\centering
\includegraphics[width=0.5\textwidth]{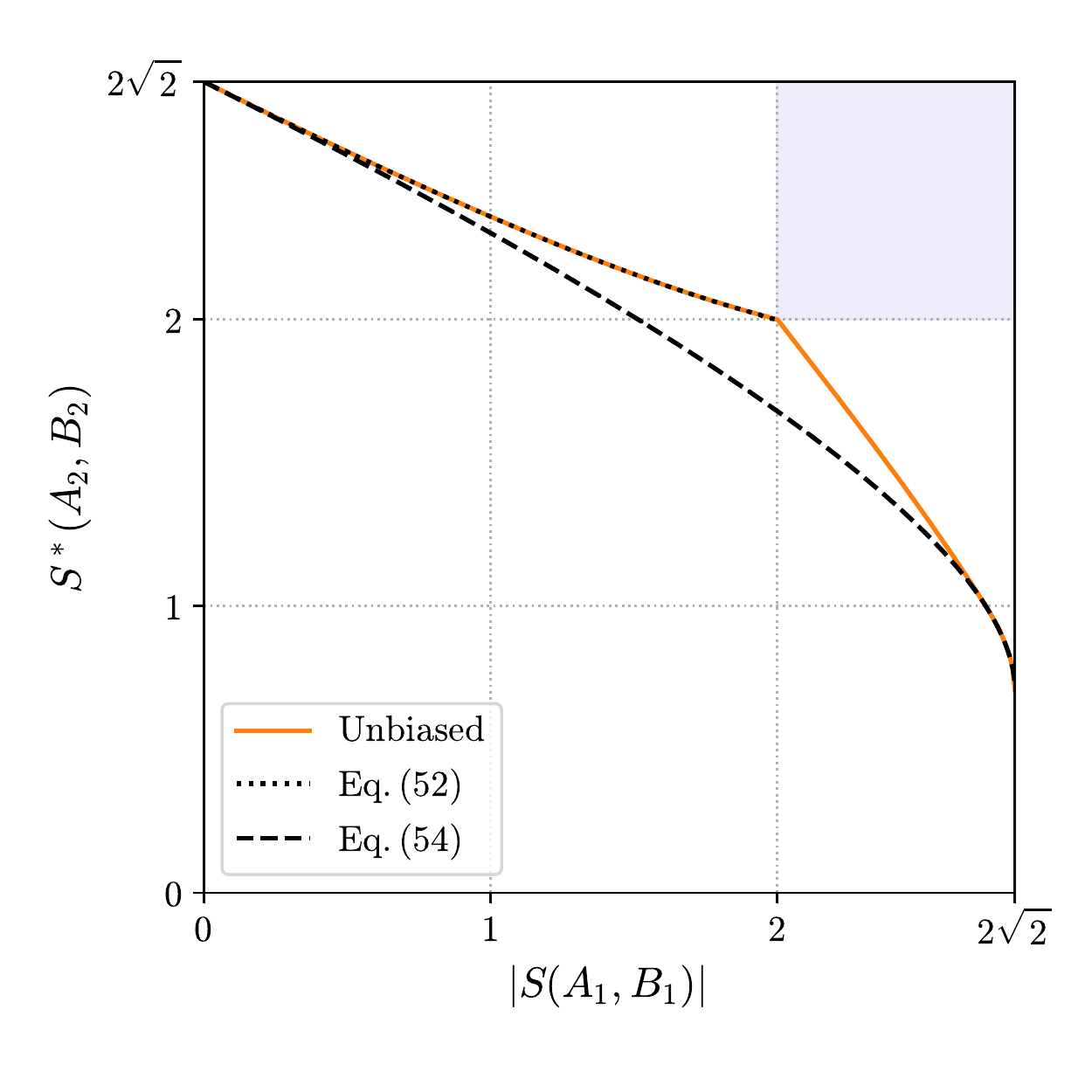}
\caption{
Semi-analytic monogamy relations for unbiased observables. The orange solid curve reproduces the numerically-generated optimal curve in Fig.~\ref{fig:monog}. The black dotted and dashed curves match the optimal curve in the regions $|S(A_1,B_1)\in[0,2]$ and $|S(A_1,B_1)\gtrsim2.72$ respectively. These curves are defined in Eqs.~\eqref{eq:case1curve} and \eqref{eq:equal_stengths_orthogonal}, found by numerically motivated ansatzes for the optimal observables measured by $A_1$ and $B_1$ (see main text). 
}
\label{fig:semianalytics}
\end{figure}

\subsubsection{First region of the optimal boundary curve}

For the first region, i.e., $|S(A_1,B_1)|\leq 2$, the numerical results suggest the ansatz $\str_{X'}=0, \str_{Y}=\str_{Y'}$ and $\bm x = \bm x' = \bm y = \bm y'=(0,1,0)^\top$ for the measurement parameters that generate corresponding section of the orange curve in Fig.~\ref{fig:monog}. This implies via Eqs.~(\ref{kdef}) and~(\ref{ldef}) that $KTL$ is diagonal, with largest singular values 
\begin{align}
s_1(KTL) &= 1 ,~
s_2(KTL) = \frac{1}{2} \left(1+\sqrt{1-\str_{X}^2}\right) \sqrt{1-\str_{Y}^2}. \label{eq:second_singular_val}
\end{align}
Now, since $S^*(A_2,B_2)$ in Eq.~\eqref{sa2b2} is an increasing function of $s_2(KTL)$, we therefore seek to find the strengths $\str_{X}$ and $\str_{Y}$ which maximize the latter. To this end, noting that $S(A_1,B_1) = 2\str_X\str_Y$ under our ansatz, we introduce the Lagrangian objective function
\begin{align}
\mathcal{L}(\str_{X},\str_{Y},\xi) &:= \frac{1}{2} \left(1+\sqrt{1-\str_{X}^2}\right)\rt{\left(1-\str_{Y}^2\right)} \nonumber\\ 
&- \xi(2\str_X\str_Y-s)
\end{align}
where $\xi$ is a Lagrange multiplier.
The stationary points of the Lagrangian, $\nabla\mathcal{L} =0$, occur when
\begin{align}
4\xi\str_Y+ \frac{\str_X\sqrt{1-\str_Y^2}}{\sqrt{1-\str_X^2}} &= 0 \\
4\xi\str_X+ \frac{\str_Y(1+\sqrt{1-\str_X^2})}{\sqrt{1-\str_Y^2}} &= 0 \\
2\str_X\str_Y &= s
\end{align}

Rewriting the first two of the Lagrangian equations in terms of the maximum reversibilities $\R_X=\sqrt{1-\str_X^2}$ and $\R_Y=\sqrt{1-\str_Y^2}$ and equating $\xi$ in these equations yields $\R_Y=\sqrt{\R_X}$, and the corresponding CHSH parameters evaluate to
\begin{align}
S(A_1,B_1) &= 2(1-\R_X)\sqrt{1+\R_X}, \\
S^*(A_2,B_2) &= \sqrt{4+(1+\R_X)^2\R_X} ,
\end{align}
in terms of $\R_X$. Plotting the parameters as $\R_X$ varies over $[0,1]$ then gives the black dotted curve in Fig.~\ref{fig:semianalytics}, which is seen to perfectly match the orange boundary curve of Fig.~\ref{fig:monog}, up to numerical error, for the region $|S(A_1,B_1)|\leq 2$. We hence propose these expressions parameterise the exact form of the boundary curve for this region. 

The Lagrangian equations can also be solved algebraically by Mathematica, to give the optimal value of $S^*(A_2,B_2)$ as an explicit function of $s:=|S(A_1,B_1)|$. 
In particular, if $h(s)$ denotes the smallest real root of the cubic polynomial
\begin{align}
	p(z) &:=   \left(s^2-4\right)z^3 - \left(3 s^2-16\right)z^2 +  \left(3 s^2-16\right)z +s^2 ,
\end{align}
then the solution for $\str_Y$ can be written as the square root of the largest root of the quadratic polynomial
\begin{equation}
	q(z) :=4  h(s) z^2 + s^2[1-h(s)]z - s^2.
\end{equation}
Denoting this solution by $\str_Y(s)$ (which we do not give explicitly in terms of $s$ here, due to its complicated form), and using $\str_X = s/(2\str_Y)$ and Eq.~\eqref{eq:second_singular_val}, we arrive at the function
\begin{equation}
	S^*(A_2,B_2) = \sqrt{ 4 + \frac{1 - \str_Y(s)^2}{4} \left(2+\sqrt{4-\frac{s^2}{\str_Y(s)^2}} \right) }
	\label{eq:case1curve}
\end{equation}
directly relating the two CHSH parameters. This function again corresponds to the black dotted line in Fig.~\ref{fig:semianalytics}, and replacing equality by $\leq$ yields our proposed optimal one-sided monogamy relation for this region.

\subsubsection{Second region of the optimal boundary curve}

For the intermediate region $2<|S(A_1,B_1)|\lesssim 2.72$, the numerical results indicate that the orange boundary curve in Fig.~\ref{fig:monog}, i.e., the solution to Eq.~\eqref{eq:optimization_problem}, occurs when $\str_{Y} = \str_{Y'}$, $B_1$'s measurement directions are determined by a single angle $\theta$ via $\bm y = (\sin\theta,\cos\theta,0)^\top, \bm y'= (-\sin\theta,\cos\theta,0)^\top$ and $A_1$'s measurement directions are determined by \emph{either} choosing $\bm x'= (1,0,0)^\top$ or $\bm x'= (\sin2\theta,\cos2\theta,0)^\top$. However, although this ansatz for the measurement parameters reduces the optimization problem to only $4$ unknown variables, $\str_X, \str_{X'}, \str_Y$ and $\theta$, we have not been able to solve it by the same methods as the previous case to obtain an explicit form for the orange boundary curve in Fig.~\ref{fig:monog} in this region. 

\subsubsection{Third region of the optimal boundary curve}

For the final region of the orange curve, $2.72 \lesssim |S(A_1,B_1)| \leq 2\sqrt{2}$, the numerics indicate that $S^*(A_2,B_2)$ obtains its extreme values when $A_1$ and $B_1$ measure observables of equal strength with orthogonal relative angles, corresponding to the ansatz $\str_X = \str_{X'} = \str_Y = \str_{Y'} =:\str$ and $\bm x = (0,1,0)^\top, \bm x' = (1,0,0)^\top, \bm y = 2^{-1/2}(1,1,0)^\top, \bm y' = 2^{-1/2}(-1,1,0)^\top$ (note these directions are the optimal CHSH directions for projective measurements~\cite{Brunner14}). Under this ansatz the two largest singular values of $KTL$ are identical, and evaluate to $\frac{1}{4} \left(\str^2-2 \left(\sqrt{1-\str^2}+1\right)\right)$, so that
\begin{equation}
S^*(A_2,B_2) = \frac{\str^2-2 \left(1+\sqrt{1-\str^2}\right)}{\sqrt{2}},
\end{equation}
which depends on only $\str$, which is uniquely determined by the constraint in Eq.~\eqref{eq:optimization_problem}, whence $S(A_1,B_1)=4\str^2/\sqrt{2}$.  Upon rearranging and substituting, we find
\begin{equation}
S^*(A_2,B_2) = \sqrt{2}-\frac{S(A_1,B_1)}{4}+\sqrt{2-\frac{S(A_1,B_1)}{\sqrt{2}}} .
\label{eq:equal_stengths_orthogonal}
\end{equation}
This is plotted as the black dashed curve in Fig.~\ref{fig:semianalytics}, and is seen to be indistinguishable from the numerically-generated orange optimal curve in Fig.~\ref{fig:monog} for this region (and is also a good approximation to the optimal curve for small values of $|S(A_1,B_1)|$). Hence, replacing equality by $\leq$ yields our proposed optimal one-sided monogamy relation for this region. Note that since this region of the orange curve matches the optimal curve for the general case of biased observables, Eq.~(\ref{eq:equal_stengths_orthogonal}) also applies to the general case.

Finally, it is straightforward to show analytically that the proposed form in  Eq.~\eqref{eq:equal_stengths_orthogonal} is indeed optimal for the case $S(A_1,B_1)=2\sqrt{2}$, i.e., that the maximum possible value of $S^*(A_2,B_2)$ for this case is $1/\sqrt{2}$. In particular, it is known that it is possible to obtain a value of $S(A_1,B_1)=2\sqrt{2}$ for a two-qubit state only if the state is maximally entangled and if projective measurements having orthogonal relative angles are made on each side~\cite{Tsirelson80,McKague12}. Hence, $s_1(T)=s_2(T)=1$, the reversibilities vanish, and $K=L={\rm diag}[\half,\half,0]$ via Eqs.~(\ref{txy}), (\ref{kdef}) and~(\ref{ldef}), and substituting in Eq.~(\ref{sa2b2}) then gives $S^*(A_2,B_2)=2[s_1(\frac14 T)^2+s_2(\frac14 T)^2]^{1/2}=1/\sqrt{2}$ as claimed. 
This analytic result also shows the conjectured one-sided monogamy relation~(\ref{unbiasedmonog}) cannot be  strengthened to the form $|S(A_1, B_1)|^d+S^*(A_2, B_2)^d\leq (2\rt{2})^d$ for $d\geq 1.76$. In particular, we need the point $(2,2)$ on or below any bounding curve (to imply the conjecture), implying the upper bound for $x^d+y^d$ can be no greater than $2^d+2^d=2^{d+1}$.  Hence, to ensure the point $(2\sqrt{2},1/\sqrt{2})$ is also on or below the curve, we require $(2\sqrt{2})^d+(1/\sqrt{2})^d\leq 2^{d+1}$, which gives $d\lesssim 1.758$.

\subsection{Analytic monogamy relations}
\label{sec:analytic}

Here we prove the one-sided monogamy relations~(\ref{monogorthog}) and~(\ref{unbiasedmonog}) for unbiased observables, for the respective cases of orthogonal directions and equal strengths on each side. We begin by showing that it is sufficient to prove them for the singlet state.

\subsubsection{Only the singlet state need be considered}

Note first, for unbiased observables, that $S(A_1,B_1)$ is linear in the spin correlation matrix $T$ of the initial state, while Eq.~(\ref{sa2b2}) can be rewritten as
\beq
S^*(A_2,B_2) = 2\| KTL\|^{(2)}_{(2)},
\eeq
where $\|M\|^{(p)}_{(q)}:=[\sum_{j=1}^q s_j(M)^p]^{1/p}$ is the singular-value matrix norm defined as per Eq.~(IV.19) of~\cite{Bhatia97}. Second, any spin correlation matrix $T$ can be represented as a convex mixture, $T=\sum_k p_k T_k$, of at most four spin correlation matrices of maximally entangled states (corresponding to the four Bell states defined by the local bases in which $T$ is diagonal)~\cite{Horodecki96}. Hence, indicating the dependence on $T$ explicitly and noting that the triangle inequality holds for absolute values and norms, we have
\begin{align}
&|S(A_1,B_1|T)|+S^*(A_2,B_2|T)\nn\\
 &~~~~~\leq \sum_k p_k\left[|S(A_1,B_1|T_k)| + S^{*}(A_2,B_2|T_k)\right]\nn\\
&~~~~~\leq \max_{T_{\rm me}} \left[|S(A_1,B_1|T_{\rm me})| + S^{*}(A_2,B_2|T_{\rm me})\right]\nn\\
&~~~~~\leq \max_{X,X',Y,Y'} \left[|S(A_1,B_1|T_0)| + S^{*}(A_2,B_2|T_0)\right] .
\label{singletmax}
\end{align}
Here, the maximum in the third line is over the spin correlation matrices of maximally entangled two-qubit states; $T_0:=-I_3$ is the spin correlation matrix of the singlet state; and the maximum in the last line is over the (compact) set of unbiased observables. The last line follows since all maximally entangled states differ from the singlet state only by local rotations, implying that maximising over a rotationally-invariant set of local observables (such as the set of unbiased observables), for a given maximally entangled state, is equivalent to maximising over the same set for the singlet state.
It follows that only the singlet state need be considered for the purposes of proving the monogamy relations, as claimed.

\subsubsection{Upper bounds for the CHSH parameters}

The final ingredients required for deriving our analytic monogamy relations are upper bounds for the CHSH parameters $S(A_1,B_1|T_0)$ and $S(A_2,B_2|T_0)$ appearing in Eq.~(\ref{singletmax}), for the case of unbiased observables.

First, for zero bias observables $X,X',Y,Y'$ with fixed strengths $\str_X, \str_{X'}, \str_Y,\str_{Y'}$ and relative measurement angles $\cos\theta=\bm x\cdot\bm x', \cos\phi=\bm y\cdot\bm y'$, measured on a singlet state, we have the tight upper bound
\begin{align} 
 |S(A_1,B_1|T_0)| \leq S_0,
 \label{iplus}
 \end{align}
with
\begin{align}
(S_0)^2&:= (\str_{X}^2+\str_{X'}^2)(\str_{Y}^2+\str_{Y'}^2) \nn\\
 &~~+2\str_X\str_{X'}(\str_{Y}^2-\str_{Y'}^2)\cos\theta \nn\\
 &~~+2\str_Y\str_{Y'}(\str_{X}^2-\str_{X'}^2)\cos\phi \nn\\
 &~~+4\str_{X}\str_{X'}\str_{Y}\str_{Y'} \sin\theta\sin\phi .
 	\label{iw}
 \end{align}}
This upper bound is proved in Appendix~\ref{appa}, and will be generalised elsewhere~\cite{Cheng21b}. Note that it simplifies to the maximum quantum value of $2\sqrt{2}$ for the case of unit strengths and orthogonal measurement directions. 

Second, from Eq.~(\ref{sa2b2}) above, noting $T_0=-I_3$, we have 
\begin{align}
 S^{*}(A_2,B_2|T_0)	&= 2\sqrt{s_1(KL)^2+s_2(KL)^2}\nn\\
& \leq 2\sqrt{s_1(K)^2s_1(L)^2+s_2(K)^2s_2(L)^2} ,
\label{bhatia}
 \end{align}
where the last line follows from Theorem~IV.2.5 of~\cite{Bhatia97}. Our strategy for obtaining analytic one-sided monogamy relations is to find upper bounds for the sums of these inequalities, that are independent of $X, X', Y,Y'$, under suitable assumptions.

\subsubsection{Monogamy for orthogonal directions on each side}

We now have the tools to prove the following result.
{\flushleft \bf Theorem 1:} 
{\it For square root measurements of arbitrary unbiased observables $X,X'$ and $Y,Y'$, made by $A_1$ and $B_1$ with equal respective probabilities, with orthogonal angles $\bm x\cdot\bm x'=0=\bm y\cdot\bm y'$ on each side, the one-sided monogamy relation 
	\beq \label{thm3}
	|S(A_1,B_1)| + S^*(A_2,B_2) \leq \frac{8\sqrt{2}}{3}  \sim 3.77
	\eeq 
	is always satisfied.}

This theorem strengthens the result proved in~\cite{Cheng21}, which required a further assumption of equal strengths on each side, and supports the Conjecture for unbiased observables in Sec.~\ref{sec:overview}. We outline its proof below, with the details left to Appendix~\ref{appb1}.

First, from Eq.~(\ref{iw}) and the orthogonality assumption,
\begin{align}
	S_0^2 &= (\str_{X}^2+\str_{X'}^2)(\str_{Y}^2+\str_{Y'}^2) +4\str_{X}\str_{X'}\str_{Y}\str_{Y'} \nn\\
	& \leq 2(\str_{X}^2+\str_{X'}^2)(\str_{Y}^2+\str_{Y'}^2) \nn\\
	&= 2(2-\R_X^2 - \R_{X'}^2) (2-\R_Y^2 - \R_{Y'}^2) , \label{thm3first}
\end{align}
where the inequality follows using $ab\leq \half(a^2+b^2)$ and the last line using the identity $\str^2= 1-\R^2$ for unbiased observables as per Eq.~(\ref{reversibility}). 

Second, again under the orthogonality assumption, the singular values of the matrices $K$ and $L$ in Eqs.~(\ref{kdef}) and~(\ref{ldef}) can be calculated (see Appendix~\ref{appb1}), and Eq.~(\ref{bhatia}) applied to give
\begin{align}
	S^*(A_2,B_2|T_0)^2&\leq \frac14 (1+\R_X)^2(1+\R_{Y})^2 \nn\\
	&~~+ \frac14 (1+\R_{X'})^2(1+\R_{Y'})^2 .
	\label{thm3second}
\end{align}

Finally, Eqs.~(\ref{singletmax}), (\ref{iplus}), (\ref{thm3first}) and~(\ref{thm3second}) may be shown to lead to  (see Appendix~\ref{appb1}) 
\begin{align}
	&|S(A_1, B_1)| + S^*(A_2, B_2)\nn \\
	&~~~\leq \max_{x, y\in [0,1]} \rt{2}(2-x^2-y^2)+\half\rt{(1+x)^4+(1+y)^4}\nn\\
	&~~~=8\sqrt{2}/3 ,
	\label{thm3third}
\end{align}
as claimed. Note that this bound coincides with the one derived in~\cite{Cheng21}, which requires the equal strength assumption. Hence, the above inequality is achieved for $\str_X=\str_Y =2\sqrt{2}/3\sim 0.943, \R_X=\R_Y=1/3$, and the optimal CHSH directions.

\subsubsection{Monogamy for equal strengths on each side}

We can also use the above tools to prove a further one-sided monogamy relation.
{\flushleft \bf Theorem 2:} 
{\it For square root measurements of arbitrary unbiased observables $X,X'$ and $Y,Y'$, made by $A_1$ and $B_1$ with equal respective probabilities, with equal strengths $\str_X=\str_{X'}$ and $\str_Y=\str_{Y'}$ on each side, the one-sided monogamy relation 
	\beq \label{thm4}
	|S(A_1,B_1)| + S^*(A_2,B_2) \leq 4
	\eeq 
	is always satisfied.}

This theorem similarly strengthens the result proved in~\cite{Cheng21}, which required a further assumption of orthogonal directions on each side, and again supports the Conjecture for unbiased observables in Sec.~\ref{sec:overview}. Its proof is outlined below, with  details given in Appendix~\ref{appb1}.

First, it follows from Eq.~(\ref{iw}) and the equal strengths assumption that
\begin{align}
	S_0^2 &= 4\str_X^2\str_Y^2 (1+ \sin\theta\,\sin\phi)\nn\\
	&\leq 4\str_{X}^2\str_{Y}^2\rt{(1+\sin^2\theta)(1+\sin^2\phi)} \nn \\
	&=4(1-\R_X^2)(1-\R_Y)^2\rt{(2-c_X)(2-c_Y)},
	\label{thm4first}
\end{align}
where the second line follows via the Schwarz inequality for the vectors $(1,\sin\theta)$, $(1,\sin\phi)$, and the third line using $\R^2=1-\str^2$ for unbiased observables as per Eq.~(\ref{reversibility}) and defining $c_X:=\cos^2 \theta$, $c_Y:=\cos^2 \phi$. 

Second, again under the equal strengths assumption, the singular values of the matrices $K$ and $L$ in Eqs.~(\ref{kdef}) and~(\ref{ldef}) can be calculated (see Appendix~\ref{appb2}), and Eq.~(\ref{bhatia}) applied to give		 
\begin{align}
	2	S(A_2,B_2|T_0)^2 
	&= \left[(1+\R_X)^2+(1-\R_X)^2 \cos^2 \theta \right] \nn\\
	&\qquad \times \left[(1+\R_Y)^2+(1-\R_Y)^2 \cos^2 \phi \right] \nn\\
	& ~~+ 4[(1-\R_X^2) |\!\cos\theta|]\,[(1-\R_Y^2)|\!\cos \phi|]  .
	\label{thm4second}
\end{align}

Finally, Eqs.~(\ref{singletmax}), (\ref{iplus}), (\ref{thm4first}) and~(\ref{thm4second}) may be shown to lead to  (see Appendix~\ref{appb2}) 
\begin{align}
&|S(A_1,B_1)|+S^*(A_2,B_2)\leq  2\max_{x, c \in [0,1]}\left[\sqrt{2-c}\,(1-x^2) \right. \nn\\
&\qquad\qquad+\left. \sqrt{ [(1+x)^2+(1-x)^2c]^2 + 4 (1-x^2)^2c}/\rt{8}\right] \nn\\
	&\qquad\qquad\qquad~~\qquad\qquad\leq 4,
	\label{thm4third}
\end{align}
as claimed in Theorem~2. Particularly, it follows from Eqs.~(\ref{thm4first}) and (\ref{thm4second}) the first inequality is achieved by $A_1$ and $B_2$ performing measurements with the same strength and relative angle, i.e., $\str_{X}=\str_{Y}$ and $\sin\theta=\sin\phi$, while the second is further saturated with $x=0$ and $c=1$, or equivalently, $\str_{X}=1$ and $\sin\theta=0$, implying that parallel directions and zero reversibilities are optimal for this case.

\subsubsection{Generalisation to a class of weak measurements}

The above analytic monogamy relations are proved for the case of square-root measurements. However, it is expected, from the argument given in Sec.~\ref{subsec4.2} (see also~\cite{Brown20}), that they also hold for arbitrary measurements, similarly to the conjectured monogamy relation~(\ref{unbiasedmonog}) for unbiased observables.  In this regard, it is worth noting that Theorems~1 and~2 indeed hold for the class of weak measurements introduced by Silva {\it et al.}~\cite{Silva15}.

In particular, for this class of measurements the reversibilities $\R_X, \R_{X'},\R_Y,\R_{Y'}$ of the post-measurement state are replaced by corresponding `quality factors' $F_X, F_{X'},F_Y,F_{Y'}$ as per Eq.~(\ref{rhof}). Further, inequalities~(\ref{thm3first}) and~(\ref{thm4first}) remain valid under this replacement, since $\str^2=1-\R^2\leq1-F^2$ as per Eq.~(\ref{frineq}), and the proofs of the theorems then follow exactly as for the case of square-root measurements.  

Likewise, noting Eq.~(\ref{rsimp}), Theorems~1 and~2 also hold for the case of unbiased observables measured as per the simple measurement protocol in Eq.~(\ref{simplemeas}). Similar generalisations apply to the one-sided monogamy relations in~\cite{Cheng21}.

\section{Qubit recycling for multiple observers}\label{Sec6. One-sided}

In this section, we use the techniques developed in Sec.~\ref{Sec4. Measurements Reversibility} to study the problem of generating Bell nonlocality between multiple pairs of independent observers. We show that this is possible for the case of multiple observers on both sides, if they share sufficiently many pairs of qubits, via a simple extension of a construction by Brown and Colbeck for the case of a single Alice and many Bobs~\cite{Brown20}. We also give an alternative extension of this construction that allows a single Alice to generate Bell nonlocality with many Bobs for a larger class of single two-qubit states.

\subsection{Arbitrarily many Alices and Bobs}
\label{sec:arbab}

Assume now that there are $M$ Alices on one side and $N$ Bobs on the other in Fig.~\ref{fig:fig1}, where each observer independently chooses between a set of two or more measurements to make on their component of a general bipartite state $\rho$. We will denote the $m$-th Alice and the $n$-th Bob by $A_m$ and $B_n$, respectively.  

It follows that if $\phi_m$ and $\chi_n$ are the CPTP maps describing the effect of local measurements made by each $A_m$ and $B_n$, on an ensemble initially described by state $\rho$, then the post-measurement state of the ensemble shared by $A_m$ and $B_n$ is given by 
\beq 
\rho''=(\phi_m\circ\dots\phi_2\circ\phi_1)\otimes (\chi_n\circ\dots \chi_2\circ\chi_1)(\rho).
\eeq
Using the notation developed in Sec.~\ref{subsec4.1}, this post-measurement state is equivalently described by the matrix
\beq
\tilde\Theta''= {\cal K}_m \dots {\cal K}_1 \Theta {\cal L}^\top_1 \dots {\cal L}_n^\top ,
\eeq
with ${\cal K}_{m\alpha\beta}:= c^{-1}\tr{\tilde\sigma_\alpha\phi_m(\tilde\sigma_\beta)}$ and ${\cal L}_{n\mu\nu}:= d^{-1}\tr{\tilde\tau_\mu\chi_n(\tilde\tau_\nu)}$, generalising Eq.~(\ref{kthetal}).
It further follows, for the case of a shared two-qubit state, that if either the local Bloch vectors vanish or the CPTP maps are unital, the corresponding spin correlation matrix $T$ is transformed to 
\beq \label{ktlgen}
T'' = K_m\dots K_1 T L_1^\top\dots L_n^{\top},
\eeq
generalising Eq.~(\ref{ktl}).

Now, the validity of the conjectures made in~\cite{Cheng21} would imply that for $M,N>1$ it is not possible for all pairs $(A_m,B_n)$ to generate Bell nonlocality if each observer is restricted to choosing between two equally-likely measurements on a single two-qubit state.  However, this does not preclude the possibility that each pair can generate Bell nonlocality  via making a greater number of measurements on higher-dimensional quantum systems, as demonstrated by the simple example below.

In particular, suppose that the observers share $M$ two-qubit states, and that  for their local component of the $q$-th two-qubit state $A_m$ and $B_n$ independently choose between equally-likely measurements of unbiased qubit observables $X_{mq}, X^\prime_{mq}$  and $Y_{nq}, Y^\prime_{nq}$, respectively. For each pair $(A_m,B_n)$ there is then a corresponding Bell inequality
\beq
S_{mn}:= \max_q\{S_q(A_m,B_n)\} \leq 2 ,
\eeq
where $S_q(A_m,B_n)=\langle X_{mq}Y_{nq}\rangle + \langle X_{mq}Y'_{nq}\rangle + \langle X'_{mq}Y_{nq}\rangle - \langle X'_{mq}Y'_{nq}\rangle$
denotes the CHSH parameter corresponding to their measurements on the $q$-th two-qubit state. For square-root measurements the local measurement operations are unital, so that the values of each $S_{mn}$ can be calculated via Eqs.~(\ref{chsh}), (\ref{ktl}) and~(\ref{ktlgen}).

To show that every one of the above Bell inequalities can be violated, with $S_{jk}>2$, we extend a construction given by Brown and Colbeck that allows a single Alice to violate the CHSH inequality with each Bob via recycling a single shared qubit state~\cite{Brown20}. The idea is to apply this construction to the $m$-th qubit pair, to ensure that $S_m(A_m,B_n)>2$ for each $n$. First, label the measured observables in the Brown-Colbeck construction by $X,X'$ for the single Alice, $A$, and by $Y_n,Y'_n$ for the $n$-th Bob, $B_n$, so that
\beq \label{brown}
S(A,B_n) >2, \qquad n=1,2,\dots,N
\eeq
 by construction. Second, choose the observables measured by $A_m$ and $B_n$ on the $q$th qubit pair to be
\beq
X_{mq}:=\left\{ \begin{matrix} X,&m=q\\ \id, & m\neq q\end{matrix}\right. , ~~~~
X'_{mq}:=\left\{ \begin{matrix} X',&m=q\\ \id, & m\neq q\end{matrix}\right. ,
\eeq
\beq
Y_{nq}=Y_n,\qquad Y'_{nq}=Y'_n .
\eeq
Since square-root measurements of the identity operator do not disturb the system, it immediately follows that
\beq S_m(A_m,B_n)=S(A,B_n) >2,
\eeq
and hence that $S_{mn}>2$ as required.

The above example, and its converse with $M$ Alices and one Bob, show that each pair can independently generate Bell nonlocality by each observer choosing between  suitable local measurements on a shared $4^{\min\{M,N\}}$ dimensional quantum system. Note that  a related example by Cabello~\cite{Cabello}, based sharing only two qubit pairs, is unsuitable in this context, as the observers do not make independent measurements (all entanglement in the first and second qubit pairs in this example is destroyed by the projective measurements made by $B_1$ and $A_1)$, respectively, implying that no later pair of observers can independently generate Bell nonlocality). 

It would be of interest to find examples requiring less measurements and/or dimensions.  For example, it is known for $M=N=2$ that each of two Alices can steer each of two Bobs, and vice versa, via recycling of a single qubit pair~\cite{Jie21}.

\subsection{Multiple Bobs}

The problem of one-sided qubit recycling, with one Alice and $N>1$ Bobs, has been well studied in previous work~\cite{Silva15,Mal16,Curchod17,Tavakoli18,Bera18,Sasmal18,Shenoy19,Das19,Saha19,Kumari19,Brown20,Maity20,Bowles20,Roy20}. In particular, Theorem~2 of~\cite{Brown20} shows that it is possible to generate CHSH Bell nonlocality between one Alice and arbitrarily many Bobs via recycling of a two-qubit state and unbiased observables, under the condition that two largest singular values of the initial spin correlation matrix $T$ satisfy 
\beq \label{browncon}
s_1(T)=1, \qquad s_2(T)>0. 
\eeq
Brown and Colbeck raised the interesting question of whether this condition was necessary as well as sufficient~\cite{Brown20}. Here we answer the simpler but related question, of whether this condition is necessary and sufficient for the case of a {\it fixed} number of Bobs. We show that it is only sufficient for this case, by constructing suitable two-qubit states with $s_1(T)<1$.

First, for a fixed number $N$ of Bobs, consider an initial two-qubit state $\rho$ satisfying the Brown-Colbeck condition~(\ref{browncon}) above, so that Eq.~(\ref{brown}) is satisfied for suitable unbiased observables $X,X'$ measured by Alice and $Y_n, Y'_n$ measured by the $n$-th Bob. Further, define the class of states 
\beq
\rho_p := p\, \rho + \frac{1-p}{4}\id\otimes\id,\qquad p \in (0, 1), \label{counterexample}
\eeq
corresponding to adding isotropic noise to $\rho$. The associated correlation matrix is then $T_p=p\,T$, with singular values 
\beq \label{tp}
{s}_1(T_p)=p\,s_1(T)=p, ~~{s}_2(T_p)=p\,s_2(T)>0.
\eeq
 Further, if Alice and each Bob choose the same measurement strategy as for $\rho$, then it follows (recalling that the observables are unbiased) that the corresponding CHSH parameters are given by
\beq
S_p(A, B_n)=p\, S(A, B_n). 
\eeq
Finally, defining
\beq
S_{\min} := \min  \{S(A, B_1), S(A, B_2), \dots, S(A, B_N)\}>2,
\eeq
where the upper bound follows from Eq.~(\ref{brown}), and
\beq
 p_{\min}:= \frac{2}{S_{\min}}<1, 
 \eeq
we have
\beq
S_p(A,B_n) \geq pS_{\min} > 2(p/p_{\min}) >2~~{\rm for}~~p\geq p_{\min}.
\eeq
Thus, Alice can violate the CHSH inequality with each of the $N$ Bobs for $p\in [p_{\min}, 1)$ in Eq.~(\ref{tp}), implying that condition~(\ref{browncon}) is not necessary, as claimed. The original question posed by Brown and Colbeck, however, as to whether  condition~(\ref{browncon}) is necessary for states suitable for sharing Bell nonlocality for {\it all} values of $N$, remains open.
	
\section{Conclusions} \label{Sec8. Conclusions}

We have studied the sequential generation of Bell nonlocality between independent observers via recycling the components of entangled systems. First, general two-valued qubit observables are characterised in~Eq.~(\ref{observable}) via the {\it outcome bias}, {\it strength}, and measurement direction, and a measurement model is provided to interpret these parameters of such observables. Based on quantum instruments, we then introduced a general formalism for measurements to describe the sequential scenarios, and review the optimal reversibility properties of square-root measurements. For measurements of a given qubit observable, the maximum reversibility and minimum decoherence are naturally defined in Eqs.~(\ref{reversibility}) and~(\ref{decoherence}) respectively. Moreover, we obtain tradeoff relations~(\ref{chain}) and~(\ref{decoherence2}) between these quantities and the strength and bias of the observable. Further, using these relations for the case of unbiased observables, we analytically obtained the strong one-sided monogamy relations in Theorems~1 and~2, as per Eqs.~(\ref{thm3}) and~(\ref{thm4}). We also provided compelling numerical evidence as displayed in Fig.~\ref{fig:monog} to support the more general conjecture in Eq.~(\ref{unbiasedmonog}) for the sequential generation of Bell nonlocality and to obtain semi-analytic results for the best possible monogamy relation as displayed in Fig.~\ref{fig:semianalytics}. Finally, we applied our tools to scenarios of arbitrary numbers of observers on one and/or two sides. We generalised the construction in~\cite{Brown20} to show that if sufficiently many pairs of entangled qubits and measurements are allowed, then arbitrarily many pairs of observers on each side can sequentially share Bell nonlocality. Moreover, a larger class of two-qubit states than in~\cite{Brown20}  was shown to allow a single Alice to share Bell nonlocality with a given number of Bobs, implying that the conditions discussed in~\cite{Brown20} are sufficient but not necessary when the number of Bobs is fixed.

There are many interesting questions left open for future work. For example, is it possible to further pin down the form of the numerically optimal orange curve in Figs.~\ref{fig:monog} and~\ref{fig:semianalytics}? Are there more efficient numerical and analytical tools to prove or disprove the one-sided monogamy conjectures in this work and Ref.~\cite{Cheng21}? Can Bell nonlocality be generated by recycling two qubits if more than two measurements are allowed per observer? (as is the case for Einstein-Podolsky-Rosen steering~\cite{Jie21}). 

It will be shown elsewhere that the bound in Eq.~(\ref{iplus}) can be extended to a generalised Horodecki criterion for nonprojective observables~\cite{Cheng21b}. Finally, we note that since our approach is based on an instrumental formalism which can incorporate the most general measurements, our analysis can be applied to similar problems in the sequential sharing of other quantum properties, such as EPR-steering and entanglement, including for cases in which more measurement settings as per observer are allowed. It would also be worth investigating if our results and methods are applicable to sequential sharing of random access codes~\cite{Mohan19,Anwer20,Foletto20b} and preparation-contextuality~\cite{Anwer21}, particularly if observer $A_1$ in such scenarios prepares states for observer $B_1$ via measurements on an entangled state (and/or vice versa).

\acknowledgements We thank Peter Brown, Ad\'an Cabello and Howard Wiseman  for helpful discussions and comments.  S. C.~is supported by the Fundamental Research Funds for the Central Universities (No.~22120210092) and the National Natural Science Foundation of China (No.~62088101). L. L.~is supported by National Natural Science Foundation of China (No.~61703254). T. J. B.~is supported by the Australian Research Council Centre of Excellence CE170100012, and acknowledges the support of the Griffith University eResearch Service \& Specialised Platforms Team and the use of the High Performance Computing Cluster ``Gowonda'' to complete this research.

\appendix

\section{Derivation of the upper bound~(\ref{iw})}
\label{appa}

First, for general qubit observables $X,X',Y,Y'$, with $X=\B_X\id+\str_X \bm\sigma\cdot\bm x$, etc., define the unit vectors
\beq \label{xdef}
\bm x_1 = \frac{\bm x+ \bm x'}{|\bm x+\bm x'|},~~~\bm x_2 = \frac{\bm x- \bm x'}{|\bm x - \bm x'|}, ~~~\bm x_3=\bm x_1\times \bm x_2,
\eeq
\beq \label{ydef}
\bm y_1 = \frac{\bm y+ \bm y'}{|\bm y + \bm y'|},~~~\bm y_2 = \frac{\bm y- \bm y'}{|\bm y - \bm y'|}, ~~~\bm y_3=\bm y_1\times \bm y_2.
\eeq
It follows that 
\beq \label{x1x2}
\bm x =\cos\frac{\theta}{2} \bm x_1+\sin \frac{\theta}{2} \bm x_2, ~~~\bm x' =\cos\frac{\theta}{2} \bm x_1-\sin \frac{\theta}{2} \bm x_2,
\eeq
\beq \label{y1y2}
\bm y =\cos\frac{\phi}{2} \bm y_1+\sin \frac{\phi}{2} \bm y_2,~~~ \bm y' =\cos\frac{\phi}{2} \bm y_1-\sin \frac{\phi}{2} \bm y_2 ,
\eeq	
where $\cos \theta=\bm x\cdot \bm x'$ and $\cos \phi=\bm y\cdot\bm y'$, i.e., $0\leq\theta\leq\pi$ is the  angle between $\bm x$ and $\bm x'$ and $0\leq\phi\leq\pi$ is the  angle between $\bm y$ and $\bm y'$. 

Second, for unbiased observables  measured on a singlet state we have $\B=0$ and $T=T_0=-I_3$, and the CHSH parameter reduces via Eqs.~(\ref{chsh}) and~(\ref{prodxy2}) to 
\begin{align}
S(A_1,B_1|T_0)&\leq -\str_X\str_{Y} \bm x \cdot\bm y - \str_X\str_{Y'} \bm x\cdot\bm y' \nn\\
& \qquad- \str_{X'}\str_{Y} \bm x' \cdot\bm y + \str_{X'}\str_{Y'} \bm x'\cdot\bm y'\nn\\
 &= -\sum_{j,k}  W_{jk} \bm x_j\cdot \bm y_k\nn\\
 &= -\tr{WR^\top},
 \label{wm}
\end{align}
\blk
where $W$ is the $3\times3$-matrix 
\beq
W:=\begin{pmatrix}
	A\cos\frac{\theta}{2}\cos\frac{\phi}{2} & B\cos\frac{\theta}{2}\sin\frac{\phi}{2} & 0\\ C\sin\frac{\theta}{2}\cos\frac{\phi}{2} & -D\sin\frac{\theta}{2}\sin\frac{\phi}{2} &0 \\ 0& 0 & 0
\end{pmatrix}
\eeq
with 
\begin{align}
	A &= \str_{X}\str_{Y}+\str_{X}\str_{Y'}+\str_{X'}\str_{Y}-\str_{X'}\str_{Y'}\nn\\
	B &= \str_{X}\str_{Y}-\str_{X}\str_{Y'}+\str_{X'}\str_{Y}+\str_{X'}\str_{Y'} \nn\\
	C &= \str_{X}\str_{Y}+\str_{X}\str_{Y'}-\str_{X'}\str_{Y}+\str_{X'}\str_{Y'} \nn\\
	D &= -\str_{X}\str_{Y}+\str_{X}\str_{Y'}+\str_{X'}\str_{Y}+\str_{X'}\str_{Y'}, \label{abcd}
\end{align}
and $R$ is the $3\times3$ matrix with coefficients
\beq
R_{jk} := \bm x_j \cdot \bm y_k .
\eeq
Note that $W$ contains information about the local measurement strengths and relative measurement directions for each side, while $R$ contains information about the relative measurement directions between the two sides. 

Third, note that
\begin{align}
(RR^\top)_{jk}&=\sum_m (\bm x_j\cdot\bm y_m) (\bm x_k\cdot\bm y_m)\nn\\
&= \bm x_j^\top \left(\sum_m \bm y_m\bm y_m^\top\right) \bm x_k =\delta_{jk},
\end{align}
using the orthonormal basis properties of $\{\bm x_j\}$ and $\{\bm y_k\}$, and so $R$ is an orthogonal matrix, i.e., a rotation or reflection (indeed, since the basis sets are right-handed by construction, $R$ is a rotation). Hence, 
\begin{align}
|S(A_1,B_1)|T_0)| \leq \max_R |\tr{WR^\top}|
\end{align}
where the maximum is over all orthogonal matrices $R$.

Fourth, suppose that $W=R'DR''$ is a singular value decomposition of $W$, for orthogonal matrices $R',R''$ and diagonal matrix $D={\rm diag}[s_1(W),s_2(W),s_3(W)]$ with singular values $s_1(W)\geq s_2(W)\geq s_3(W)\geq0$. Substitution then gives
\begin{align}
	|S(A1,B1)| &\leq \max_R |\tr{DR''R^TR'}| \nn\\
	&= \max_{\tilde R} |\tr{D\tilde R}| \nn\\
	&= \max_{\tilde R} |\sum_j s_j(W)	\tilde{\bm x}_j\cdot\tilde{\bm y}_j| \nn\\
	&\leq  \max_{\tilde R} \sum_j s_j(W) |\tilde{\bm x}_j\cdot\tilde{\bm y}_j| \nn\\
	&\leq S_0:=\sum_j s_j(W)
\end{align}
where $\tilde R:=R''R^TR'$ is an orthogonal matrix, implying there are local coordinate systems $\{\tilde{\bm x}_j\}$ and $\{\tilde{\bm y}_j\}$ such that $\tilde{R}=\tilde{\bm x}_j\cdot\tilde{\bm y}_k$, and we have used 
$\tilde{\bm x}_j\cdot\tilde{\bm y}_j\leq 1$ with equality for $\tilde{\bm x}_j\equiv \tilde{\bm y}_j$. Note that the upper bound is achievable by construction.

Finally, to show that $S_0$ above has the explicit formula given in Eq.~(\ref{iw}) of the main text, let $\tilde W$ denote the upper $2\times2$ submatrix of $W$, and $w_\pm$ denote the eigenvalues of $\tilde W^\top \tilde W$ (i.e., the nonzero eigenvalues of $W^\top W$). It follows that $S_0=\sqrt{w_+}+\sqrt{w-}$.
The identities
\beq 
w_+ + w_- = \tr{\tilde W^\top \tilde W},~ w_+w_-=\det(\tilde W^\top \tilde W)=\det(\tilde W)^2, \nn
\eeq
then imply that
\begin{align}
S_0^2&=
	(\sqrt{w_+}+\sqrt{w_-})^2 \nn\\
	&= w_++w_- + 2\sqrt{w_+w_-} \nn\\
	&=  \tr{\tilde W^\top \tilde W} + 2 |\det(\tilde W)| .
	\label{cor2proof}
\end{align}
Explicit calculation of the trace and determinant yields Eq.~(\ref{iw}), as desired.

\section{Derivation of one-sided monogamy relations}
\label{appb}

\subsection{Proof of Theorem 1}
\label{appb1}

First, as already noted in Eq.~(\ref{thm3first}) of the main text, it follows from Eq.~(\ref{iw}) and the orthogonality assumption $\bm x\cdot\bm x'=0=\bm y\cdot\bm y'$ that
\begin{align}
	S_0^2 &= (\str_{X}^2+\str_{X'}^2)(\str_{Y}^2+\str_{Y'}^2) +4\str_{X}\str_{X'}\str_{Y}\str_{Y'} \nn\\
	& \leq 2(\str_{X}^2+\str_{X'}^2)(\str_{Y}^2+\str_{Y'}^2) \nn\\
	&= 2(2-\R_X^2 - \R_{X'}^2) (2-\R_Y^2 - \R_{Y'}^2) . \label{Sa1b1}
\end{align}

Further, again using the orthogonality assumption, one has $I_3=\bm x\bm x^\top+\bm x'\bm x'^\top + \bm x''\bm x''^\top$, with $x'':=\bm x\times\bm x'$, and Eq.~(\ref{kdef}) for the matrix $K$ simplifies to
\begin{align}
	K&=\frac{1+\R_{X'}}{2} \bm x\bm x^\top + \frac{1+\R_{X}}{2} \bm x'\bm x'^\top + \frac{\R_X+\R_{X'}}{2}\bm x''\bm x''^\top . \nn
\end{align}
Hence, assuming $\R_{X'}\leq \R_X$ without any loss of generality, the first two singular values of $K$ can be directly read off as
\beq
s_1(K)=\half(1+\R_X),~~s_2(K)=\half(1+\R_{X'}).
\eeq
Similarly,  assuming $\R_{Y'}\leq \R_Y$ without any loss of generality, we find via Eq.~(\ref{ldef}) that
\beq
s_1(L)=\half(1+\R_Y),~~s_2(L)=\half(1+\R_{Y'}) .
\eeq
Using Eq.~(\ref{bhatia}) then gives
\begin{align}
	S^*(A_2,B_2|T_0)^2&\leq \frac14 (1+\R_X)^2(1+\R_{Y})^2 \nn\\
	&~~+ \frac14 (1+\R_{X'})^2(1+\R_{Y'})^2 \nn\\
	&\leq \frac14\sqrt{(1+\R_X)^4 +(1+\R_{X'})^4} \nn\\
	&~~\times\sqrt{(1+\R_Y)^4 +(1+\R_{Y'})^4} , \label{Sa2b2}
\end{align}
using $\bm a\cdot \bm b\leq |\bm a||\bm b|$ for $\bm a=((1+\R_X)^2,(1+\R_{X'}^2))$, etc.

Now, defining the functions
\begin{align}
	g_1(x,y) &:= 2^{1/4}\rt{2-x^2 - y^2}, \nn \\
	g_2(x,y) &:= 2^{-1/2}[(1+x)^4 +(1+y)^4]^{1/4}, \nn 
\end{align}
for $x,y\in[0,1]$, it immediately follows from Eqs.~(\ref{iplus}), (\ref{Sa1b1}) and~(\ref{Sa2b2}) that
\begin{align}
	&|S(A_1, B_1)|T_0| + S^*(A_2, B_2|T_0)\nn \\
	&\leq g_1(\R_X, \R_{X'}) g_1(\R_Y, \R_{Y'})+g_2(\R_X, \R_{X'})g_2(\R_Y, \R_{Y'}) \nn \\
	&\leq \rt{g_1(\R_X, \R_{X'})^2+g_2(\R_X, \R_{X'})^2} \nn\\
	&~~~\times \rt{g_1(\R_Y, \R_{Y'})^2+g_2(\R_Y, \R_{Y'})^2} \nn \\
	& \leq \max_{x, y\in [0,1]} \left\{ g_1(x, y)^2+g_2(x, y)^2 \right\} \nn \\
	&= \max_{x, y\in [0,1]} G(x, y), \label{Gxy}
\end{align}
where the third line follows from $\bm a\cdot \bm b\leq |\bm a||\bm b|$ for $\bm a=(g_1(\R_X,\R_{X'}),g_2(\R_X,\R_{X'}))$, etc., and
\beq
G(x,y):=\rt{2}(2-x^2-y^2)+\half\rt{(1+x)^4+(1+y)^4} 
\eeq
as per the second line of Eq.~(\ref{thm3third}) of the main text.

Solving $\partial G/\partial x=0=\partial G/\partial y$ yields the conditions
\beq \label{gxgy}
g(x)=g(y) = \frac{1}{2\rt{2}\rt{(1+x)^4+(1+y)^4}},
\eeq
for a local extremum of $G(x,y)$, where
\beq
g(x):= \frac{x}{(1+x)^3}.
\eeq
Now, it is easy to check that $g(x)=g(y)$ has one solution, $y=x$, for $x\leq x_0:=\sqrt{5}-2$; two solutions, $y=x$ and $y=x^*$, for $x_0<x\neq\half$; and one solution, $y=x=\half$, for $x=\half$ (corresponding to the maximum of $g(x)$). For the solution $y=x^*$,  conditions~(\ref{gxgy}) simplify to
\beq
\frac{x}{(1+x)^3} =  \frac{1}{2\rt{2}\rt{(1+x)^4+(1+x^*)^4}} ,
\eeq
yielding
\beq
x^*=\frac{(x+1)\sqrt{x(x+4)}-x(x+3)}{2x} ,
\eeq
However, substituting this into the right hand side of Eq.~(\ref{gxgy}), and plotting both sides over the range $[x_0,1]$ leads to $y=x^*=\half$, and hence via $g(x)=g(y)$ that $x=y=\half$. Hence only the universal solution $y=x$ can generate local extrema. But for this solution, the above conditions simplify to $1+x=4x$, yielding a local maximum value of $G$ at $x=y=\frac13$, with
\beq \label{Gthird}
G(1/3,1/3) = \frac{8\sqrt{2}}{3} .
\eeq
To show that this is the global maximum of $G(x,y)$, one needs to check its values on the boundaries of the domain. One finds that $f(x,1)=f(1,x)\leq 3.49< G(\frac13,\frac13)$ and $f(x,0)=f(0,x)\leq 3.71<G(\frac13,\frac13)$ for $x\in [0,1]$.  Hence, $G(\frac13,\frac13)$ is indeed the  global maximum.

Hence, combining Eqs.~(\ref{singletmax}), (\ref{Gxy}) and~(\ref{Gthird}), we have the additive monogamy relation
\begin{align}
	|S(A_1,B_1)| + S^*(A_2,B_2) \leq  \frac{8\rt{2}}{3} 
\end{align}
for unbiased observables with orthogonal measurement directions on each side, as claimed in Theorem~1.

\subsection{Proof of Theorem 2}
\label{appb2}

First, as per Eq.~(\ref{thm4first}) of the main text, it follows from Eq.~(\ref{iw}) and the equal strengths assumption that
\begin{align}
	S_0^2 &= 4\str_X^2\str_Y^2 (1+ \sin\theta\,\sin\phi)\nn\\
	&\leq 4\str_{X}^2\str_{Y}^2\rt{(1+\sin^2\theta)(1+\sin^2\phi)} \nn \\
	&=4(1-\R_X^2)(1-\R_Y)^2\rt{(2-c_X)(2-c_Y)} .
\end{align}
Hence, taking square roots and recalling that the geometric mean is never greater than the arithmetic mean, 
\begin{align}
	S_0 &\leq 2\sqrt{\sqrt{2-c_X}(1-\R_X^2)\rt{2-c_Y}(1-\R_Y)^2} \nn\\
	&\leq \sqrt{2-c_X}\, [1-\R_X^2]+\rt{2-c_Y}[1-\R_Y^2] .
	\label{iwbound}
\end{align}

Second, substituting~Eq.~(\ref{x1x2}) of Appendix~\ref{appa} into Eq.~(\ref{kdef}),  we have
\beq
K=\R_XI_3+(1-\R_X)(\cos^2\frac{\theta}{2}\bm x_1\bm x_1^\top + \sin^2\frac{\phi}{2}\bm x_2\bm x_2^\top)
\eeq
for equal reversibilities, in terms of the orthogonal unit vectors $\bm x_1$ and $\bm x_2$. Since $K$ is a symmetric matrix, this immediately allows us to read off the two largest singular values of $K$ as the corresponding two largest eigenvalues of $K$,
\beq
\lambda_\pm(K) =\half (1+\R_X)\pm \half(1-\R_X)|\cos\theta|.
\eeq
Similarly, the two largest singular values of $L$ follow from Eqs.~(\ref{ldef}) and~(\ref{y1y2}) as
\beq
\lambda_\pm(L) =\half (1+\R_Y)\pm \half(1-\R_Y)|\cos\phi|.
\eeq
Hence, using $(a+b)(c+d)+(a-b)(c-d)=2(ac+bd)$, the bound in Eq.~(\ref{bhatia}) simplifies to 
\begin{align}
	2	S(A_2,B_2|T_0)^2 &\leq 2\lambda_+(K)^2\lambda_+(L)^2 + 2\lambda_-(K)^2\lambda_-(L)^2 \nn\\
	&= \left[(1+\R_X)^2+(1-\R_X)^2 \cos^2 \theta \right] \nn\\
	&\qquad \times \left[(1+\R_Y)^2+(1-\R_Y)^2 \cos^2 \phi \right] \nn\\
	& ~~+ 4[(1-\R_X^2) |\!\cos\theta|]\,[(1-\R_Y^2)|\!\cos \phi|] \nn\\
	&= PQ +4MN = (P,2M)\cdot (Q,2N) \nn\\
	&\leq \sqrt{P^2+4M^2}\sqrt{Q^2+4N^2} \nn\\
	&=f(\R_X,c_X)^2 f(\R_Y,c_Y)^2 \nn\\
	&\leq \frac14 [f(\R_X,c_X)^2+f(\R_Y,c_Y)^2]^2
	\label{relaxedS2}
\end{align}
where
\beq
f(x,c)^4:= [(1+x)^2+(1-x)^2c]^2 + 4 (1-x^2)^2c .
\eeq

Finally, combining Eqs.~(\ref{iwbound}) and~(\ref{relaxedS2}) gives

\begin{align}
	S_0+S^*(A_2,B_2)&\leq  g(\R_X,c_X) + g(\R_Y,c_Y)\nn\\
	&\leq  2\,\max_{x, c \in [0,1]}g(x, c),
\end{align}
where the right hand side corresponds to the first upper bound in Eq.~(\ref{thm4third}), with
\beq
g(x,c):= \sqrt{2-c}\,(1-x^2) + f(x,c)^2/\sqrt{8}.
\eeq
Numerically maximising $g(x,c)$ over $x$ and $c$ gives $g_{\max}=2$, corresponding to $x=0$ and $c=1$. This implies that parallel directions and zero reversibilities are optimal for this case, and yields, via Eqs.~(\ref{singletmax}) and~(\ref{iplus}), the one-sided monogamy relation
\beq
|S(A_1,B_1)|+|S(A_2,B_2)| \leq  4 
\eeq
for unbiased observables with equal strengths on each side, as claimed in Theorem~2.

\end{document}